\documentclass{aa}

\usepackage[varg]{txfonts}

\usepackage{natbib}
\usepackage{subfig}
\usepackage{float}
\bibpunct{(}{)}{;}{a}{}{,} 

\usepackage{ulem}
\usepackage{hyperref}

\newcommand{\hi}{\textsc{H\,i}}
\newcommand{\hb}{H$\beta$}
\newcommand{\ha}{H$\alpha$}

\newcommand{\hei}{He\,\textsc{i}}

\newcommand{\sii}{\textsc{[S\,ii]}}

\newcommand{\oi}{[O\,\textsc{i}]}

\newcommand{\nii}{\textsc{[N\,ii]}}

\newcommand{\oiii}{\textsc{[O\,iii]}}

\begin{document}

\titlerunning{Mapping the DIBs $\lambda$5780 and $\lambda$6284 in the LIRG merger \object{NGC 6240}}
\title{Mapping the diffuse interstellar bands $\lambda$5780 and $\lambda$6284 in the luminous infrared galaxy merger \object{NGC 6240}}

\author{C.D. van Erp\inst{1} \and A. Monreal-Ibero\inst{1} \and J.C. Stroo\inst{1} \and P.~M.\ Weilbacher\inst{2} \and J.V. Smoker \inst{3}}

\institute{Leiden Observatory, Leiden University, P.O. Box 9513, 2300 RA Leiden, The Netherlands \and Leibniz-Institut für Astrophysik Potsdam (AIP), An der Sternwarte 16, 14482 Potsdam, Germany \and European Southern Observatory, Alonso de Cordova 3107, Vitacura, Santiago, Chile}

\date{\today}

\abstract{Diffuse interstellar bands (DIBs) are faint absorption features of a generally unknown origin. Observational constraints on their carriers have been provided in the vast majority of the cases thanks to observations in our Galaxy. Detections in other galaxies are scarce, both in the Local Group and beyond. However, they can further constrain the nature of the carriers by sampling different environments. They can put the ubiquity of the molecules creating these features to the test.}
{We aim to map some of the strongest DIBs in an environment that has not been tested thus far: a system harbouring two active galactic nuclei (AGNs). We  explore the relation of these DIBs with other components and properties of the interstellar medium, in particular, the dust traced by the attenuation, the sodium absorption doublet, and previously published maps of the atomic and molecular matter.
}
{We used archival Multi Unit Spectroscopic Explorer (MUSE) data of the luminous infrared galaxy (LIRG) \object{NGC 6240}. We spatially binned the data with the Voronoi binning technique and
modeled the emission of the underlying stellar population with the \texttt{pPXF} code.  
We measured the spectral features of interest, both in the emission and absorption, with a self-written algorithm using multiple Gaussians.}
{We mapped the DIB$\lambda$5780 over an almost contiguous area of $\sim$76.96 kpc$^{2}$ in the center of the system. We also traced the DIB$\lambda$6284 over two separate areas toward the north and south of the system, with an extent of $\sim$21.22 kpc$^2$ and $\sim$31.41 kpc$^2$ (with a total detected area of $\sim$59.78 kpc$^2$). 
This is the first time that the $\lambda$6284 DIB has been  mapped outside our Galaxy. Both maps were compared with the attenuation on the overall stellar population and the ionized gas.
As expected, both DIBs are detected in locations with high attenuation ($E(B-V)_\mathrm{Gas} \gtrsim 0.3$ and $E(B-V)_\mathrm{Stellar} \gtrsim 0.1$), supporting the connection between DIB carriers and dust. Moreover, when compared with other galaxies, DIBs are better correlated with the stellar (rather than the ionized gas) attenuation. In particular, the DIB$\lambda$6284 presents a stronger correlation with reddening than the $\lambda$5780 DIB, as determined by the Pearson correlation coefficient with value $\rho_{t,\lambda 6284} = 0.82$ and $\rho_{t,\lambda 5780} = 0.77$. This better correlation can be attributed to a different nature of the carriers causing these DIBs or a combined effect of a dependence with the metallicity and the different locations where these DIBs have been measured. We argue that the latter effect can have a more substantial impact as both $\lambda$5780 and $\lambda$6284 DIBs belong to the $\sigma$-DIB family; thus, they are expected to have similar properties.
In addition, we show that \ion{Na}{i} D is strongly correlated with both DIBs. We advocate for the utilization of DIBs as a first-order tracer of  specific amounts of material in cases where \ion{Na}{i} D reaches saturation. This saturation effect can be an observational complication in systems with a large amount of gas, such as (U)LIRGs.}
{The findings presented here show that DIB carriers can exist and survive in an environment as extreme as a galaxy hosting an AGN. These features enable us to envision the possibilities of integral field spectrographs in studying DIBs well beyond our Galaxy.}


\maketitle

\begin{figure*}[th]
    \centering
    \includegraphics[width=17cm, angle=0, trim={1cm 1cm 2cm 1cm}, clip]{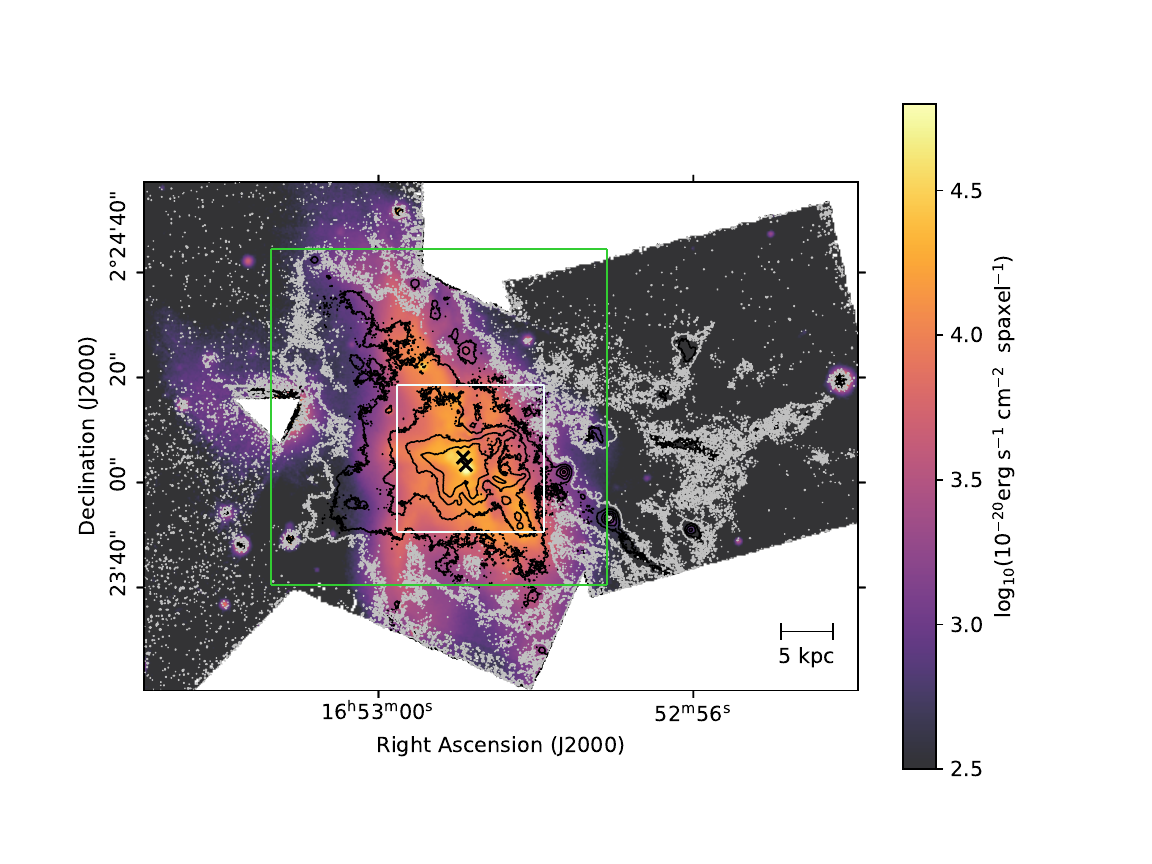}
    \caption{Line-free continuum map created from the full MUSE datacube in the observed frame with a wavelength range of $\lambda 5700~\AA- \lambda5850~\AA$. In this figure, north is to the top and east to the left. The white contour is the peak H$\alpha$ flux with a logarithmic intensity of 2, clearly showing an extended emission of H$\alpha$ to the west of this merger. The black contours  also trace the integrated H$\alpha$ flux in logarithmic intensities [2.5,3.,3.5,4.] with units of f(H$\alpha$)=$10^{-20}$ erg s$^{-1}$ cm$^{-2}$ spaxel$^{-1}$, which are also used throughout the rest of this paper. The green square indicates the region of this merger that is used for nearly all spatial maps in this paper, for instance, in Fig. \ref{fig:zoomcube}. The white square is a zoomed-in section around the nucleus, used in Fig. \ref{fig:CentreDib}. The black crosses show the locations of the AGNs \citep{Komossa2003}. 
    }
    \label{fig:datacube}
\end{figure*}

\section{Introduction}
Diffuse interstellar bands (DIBs) are faint absorption features observed in the spectra emitted by astronomical sources. They are most often observed in lines of sight toward stars, but also toward galaxies and quasars \citep[e.g.,][]{Wu2022, MonrealIbero2018, Lan2015}. \citet{Heger1922} detected the first two DIBs. Later, \citet{Merrill1936} showed them to be of interstellar origin and not an intrinsic property of a stellar absorption spectrum. There are many DIBs, with \citet{Fan2019} identifying at least 559, most of them lying in the optical domain \citep[e.g.,][]{Herbig1995,Galazutdinov2000,Hobbs2009}. More recently, several DIBs have also been identified in the near infrared \citep{Geballe2011,Cox2014,Elyajouri2017,Hamano2022, Smoker2023}. In addition, a few DIBs have tentatively been identified in the UV \citep{Destree2009, Bhatt2015}.

Nearly all of these DIBs are of an unknown origin. Four have been identified as being the absorption feature of the $\mathrm{C_{60}}^+$ molecule \citep{Campbell2015, Spieler2017,  Cordiner2019, Nie2022}. However, there is no consensus on this identification
(see: \citealp{Galazutdinov2017}; for a rebuttal: \citealp{Lallement2018}).
Different varieties of carbonated molecules are among the possible candidates that are responsible for these bands. These molecules are named the carriers.
This includes, for example,
hydrocarbon chains \citep[e.g.,][]{Maier2004}, polycyclic aromatic hydrocarbons \citep[PAHs, e.g.,][]{Salama1996}, or fullerenes \citep[e.g..][]{Iglesias2019}.
These are all molecules with many atoms, implying an immense parameter space to be constrained, making it very difficult to identify the exact carrier of each DIB. For example, there are over 1.2 million PAHs with 100 or fewer carbon atoms \citep{Cami2018}.
In that sense, their detections in a large variety of astronomical environments can narrow down the search. Studies of some DIBs in our Galaxy suggest that they are correlated with extinction and, consequently, with the amount of dust \citep[e.g.,][]{Lan2015, Baron2014, Puspitarini2013}. That said, the absence of polarization features in DIBs imply that the carriers themselves are unlikely to be dust \citep{Cox2011}.
They have also been associated with several phases of the interstellar medium (ISM). For instance, all of them clearly correlate with the \hi{} column density \citep{Herbig1993, Friedman2011, Lan2015}. On the contrary, the relation between DIB strength and molecular $\mathrm{H}_2$ column density is not so straightforward, as not all DIBs present a clear correlation. Some DIBs even present an anti-correlation \citep{Friedman2011, Welty2014, Lan2015}.
Moreover, different DIBs react differently to the ambient UV radiation field. Specifically,  the equivalent width ratio of the DIBs EW($\lambda$5797)/EW($\lambda$5780) has been proposed to be a tracer of this UV radiation field \citep{Sollerman2005,Vos2011, Ensor2017}. DIBs seem to be strongly related with colder dense interstellar clouds, as traced by \ion{Na}{i} D, as well as, albeit less strong, with warmer diffuse interstellar clouds, as traced by CaII K \citep{Kos2013, Vanloon2013, Bailey2015, Farhang2015}. 
The broad variety of DIBs support the need of several carriers to explain the whole set of detected bands. In fact, the only pair of DIBs that have been determined so far to correlate close to perfectly with each other are the DIB pairs $\lambda$6196.0-$\lambda$6613.6 \citep{McCall2010} and $\lambda$6521-$\lambda$6613.6 \citep{Ebenbichler2024}. DIBs seem to relate to each other by groups, forming the so-called ''families,'' suggesting that their carriers may share some properties. 
A recent study by \citet{Fan2022} identified four distinguishable groups out of 54 different DIBs, showing their preferred interstellar environment and their correlation regarding each other.
These relations between DIBs and other phases and properties of the ISM are certainly useful for putting constraints on the nature of the carriers. Moreover, these relations offer the possibility to use the DIBs as diagnostic tools to learn about the ISM where these DIBs are located. However, these are relations derived using information obtained mainly in one galaxy, the Milky Way, and thus only a restricted range of environmental conditions have been probed. It would be desirable to learn to which extent these relations remain valid, when the conditions differ drastically (e.g., a different metallicity, a harder ionization field, etc.). On an even more basic level, we do not know yet whether DIB carriers are omnipresent or only found at specific conditions (i.e., those sampled so far).
The way to move forward regarding both ubiquity and relation with other phases of the ISM, is exploring the properties of DIBs outside of our Galaxy. Given the faintness of these features, this is extremely challenging. However, current facilities have allowed the community to already make the first steps. 
Outside of the Milky Way, DIBs have been observed in the Small and Large Magellanic clouds (SMC and LMC, respectively), where \citet{Cox2006} managed to detect twelve DIBs in five lines of sight, and \citet{Welty2006} detected DIBs $\lambda$5780, $\lambda$5797 and $\lambda$6284 toward 20 stars. More recently, \citet{Bailey2015} mapped the DIBs at $\lambda$5780 and at $\lambda$5797 in the SMC and LMC by using 666 different spectra of early-type stars.
Likewise, \citet{Cordiner2011} detected the DIBs at $\lambda$5780 and at $\lambda$6284  in the Andromeda Galaxy (\object{M{}31}) in eleven lines of sight.
Outside of the Local Group, several DIBs have been observed toward very bright sources. For instance, \citet{Sollerman2005} detected a total of thirteen DIBs toward two dusty supernovae. Building further, \citet{Phillips2013} detected DIB$\lambda$5780 toward 32 type Ia supernovae. 
\citet{Lawton2008} directly measured the equivalent width of six DIBs ($\lambda$4428, $\lambda$5705, $\lambda$5780,
$\lambda$5797, $\lambda$6284, and $\lambda$6613) in a total of seven damped Lyman-$\alpha$ absorbers.
Finally, \citet{Ritchey2015} detected eight DIBs, ($\lambda$5780, $\lambda$5797, $\lambda$5850, $\lambda$6196, $\lambda$6204, $\lambda$6376, $\lambda$6379, and $\lambda$6614) in the nucleus of galaxy \object{NGC 5195}. 

\begin{table*} 
\caption{Details of the observations}\label{tab:obs}
\begin{tabular}{l ll cc rr l}
Night      & Program/PI          & Field$^b$ & PA         & Depth         & \multicolumn{2}{c}{Seeing} & Comments \\
(UT)$^a$   &                     &           & [$\degr$]  & [s]           & AG$^c$    & MUSE$^d$       &          \\
\hline\hline
2015-06-10 & 095.B-0482/Treister & C         & 65         & 2$\times$1200 & 0\farcs98 & 0\farcs83      & plus 400{}s sky field \\
2015-06-11 & 095.B-0482/Treister & C         & 65         & 2$\times$1200 & 0\farcs85 & 0\farcs75      & plus 400{}s sky field \\
\hline
2016-07-13 & 097.B-0588/Privon   & E         & 45,315,225 & 3$\times$965  & 1\farcs06 & 0\farcs79      & \\
\hline
2017-07-30 & 099.B-0456/Privon   & NE        & 0,90,180   & 5$\times$965  & 0\farcs87 & 0\farcs86      & exp.\ 2, 3, and 5 unusable$^e$ \\
2017-08-16 & 099.B-0456/Privon   & NE        & 0,90,270   & 3$\times$960  & 0\farcs95 & 1\farcs04      & \\
\hline
2017-08-16 & 099.B-0456/Privon   & W         & 15,105,195 & 3$\times$965  & 0\farcs85 & 0\farcs88      & \\
\hline
\end{tabular}\\
{\footnotesize
$^a$ This column gives the UTC date of the start of the night.\\
$^b$ \textbf{C}entral, \textbf{E}astern, \textbf{N}orth-\textbf{E}astern, and
     \textbf{W}estern fields were observed.\\
$^c$ Measured using Gaussian fits by the VLT autoguiding system, averaged over
     each exposure.\\
$^d$ Simple FWHM estimate from all auto-detected point-like objects at the
     central wavelength in the MUSE cube.\\
$^e$ The derotator failed during exposures 2 and 3, causing elongated features,
     clouds started appearing during exposure 5.
}
\end{table*}

All the aforementioned examples of DIB detections made use of one or a few spectra along a specific line of sight. 
An alternative possibility would be co-adding the signal of many spectra, which makes the DIB detection easier as it increases the signal-to-noise ratio (S/N).
This can easily been done with multiplexed spectrographs, that collect many spectra at once. In particular, those based on integral field spectroscopy (IFS) map a continuous area of the sky, allowing for a trade-off between the desired S/N and spatial resolution.
The advent of the Multi Unit Spectroscopic Explorer \citep[MUSE,][]{Bacon+10}, the most efficient IFS-based instrument currently available, makes it possible to apply this strategy to detect (and map) spectral features as faint as the DIBs at an unprecedented spatial resolution.
This idea was first presented in a study with MUSE on the \object{AM 1353-272} galaxy interacting system, with the observation of a radially decreasing gradient for the DIB$\lambda$5780  \citep{MonrealIbero2015}.
The potential of this philosophy is better appreciated in \citet{MonrealIbero2018}, where   the DIBs were mapped at $\lambda$5780 and $\lambda$5797 over the \object{Antennae Galaxy}. In this study, the DIBs in relation with other phases and properties of the ISM in the system, namely, the reddening, atomic hydrogen, molecular gas, and the mid infrared emission at 8 $\mu$m, was also explored.
In that work, no attempt was made to detect and map the much stronger DIB at $\lambda$6284, since at the redshift of the galaxy, this DIB falls in a spectral range affected by telluric features.
Continuing the series of experiments initiated by \citet{MonrealIbero2015}, here we aim to detect and map some of the strongest DIBs in another system with high dust content: the advanced merger \object{NGC 6240}.

\object{NGC 6240} is a well known system hosting two active galactic nuclei \citep[AGNs,][]{Komossa2003}. 
For this advanced merger, the time for merging of the supermassive black holes is estimated to be less than $\sim55~\mathrm{Myr}$ \citep{Sobolenko2021}.
Recently, a study by \citet{Kollatschny2020} implied the existence of a third non-active black hole. However \citet{Treister2020} did not find evidence for this third black hole in their high-resolution ALMA maps of the CO lines and the dust continuum.
To the east and west of the AGNs there are strong ionized outflows of material \citep{Muller2018} and there have been multiple starbursts in the past 100 million years, the last of which is still ongoing \citep{Lutz2003, Yoshida2016}. \object{NGC 6240} has a redshift of z = 0.024323 \citep{Iwasawa1998}, which implies a shift in interstellar lines and bands of $\sim$ 7300 km/s and $\sim$150 \AA\ with respect to observations at z=0. An overall visualization of the system is presented in Fig.~\ref{fig:datacube}.

With a far-infrared (FIR) luminosity of $\log (L_{IR}/L_\odot)$= 11.93 \citep{Kim2013}, \object{NGC 6240} is classified as a luminous infrared galaxy (LIRG). As such, it has large amounts of dust and therefore also heavy attenuation \citep{Armus2009}. This makes \object{NGC 6240} a good target to search for DIBs at distances and conditions not explored so far.
In fact, \citet{Heckman2000} managed to detect the DIB$\lambda$5780 with an equivalent width of $640 \pm 70 ~\mathrm{\AA}$.
In this paper, we aim at detecting and mapping some of the strongest DIBs: $\lambda$5780, $\lambda$5797, and $\lambda$6284. We  test the relation between these DIBs and other components of the ISM. In particular, we  link them with dust, atomic and molecular clouds, and metallicity.

The paper is laid out as follows. Section \ref{sec:data} describes the data used in the study. In Sect. \ref{sec:method}, we explained the methods we used for processing  the data. In Sect. \ref{sec:results}, we present our results and discuss them in Sect. \ref{sec:discussion}. In Sect. \ref{sec:conclusion}, the conclusions that can be drawn from our results are presented. Throughout this paper, we assume a Hubble parameter value of $H_0$ = 70 km/s/Mpc,  resulting in a distance to \object{NGC 6240} of 104 Mpc and corresponding to a scale of 504.2 pc/$^{\prime\prime}$.

\section{The data}\label{sec:data}

NGC\,6240 was observed with the MUSE instrument \citep{Bacon+10} installed at
the fourth VLT unit telescope (``Yepun'') in the wide-field mode in
seeing-limited operations (WFM-NOAO-N) at four different positions around its
center. The details of the exposures are given in Table~\ref{tab:obs}, the sky
conditions for all exposures that were finally used were clear or photometric.
In this mode, MUSE has a usable wavelength range of 4750--9350~\AA{} and a
sampling of 0\farcs2$\times$0\farcs2$\times$1.25\,\AA{} over a field of about
$1\arcmin\times1\arcmin$, with an average spectral resolving power of $R\sim2700$ or $\sim$111 km\,s$^{-1}$.

The data were downloaded in raw format for the science data itself and for the
standard stars, all other calibrations were retrieved in processed form as
associated with the corresponding raw exposure by the ESO archive system
\citep{2018SPIE10704E..16R}. The reduction was carried out with the standard
MUSE pipeline \citep[v2.8.5,][]{musepipeline}. We ran the basic processing of
the science and standard star exposures as well as the spectral response
computation using default parameters. The algorithm included bias subtraction,
flat-fielding, spectral tracing, wavelength calibration, geometrical
calibration, illumination correction, and a twilight-sky flat-field correction. No telluric correction was applied. 

To estimate the contamination due to atmospheric water in the observations, we used {\sc skycalc}\footnote{The ESO sky model (\url{https://www.eso.org/observing/etc/skycalc/}); see \citet{Noll2012} and \citet{Jones2013}.} to create a synthetic spectrum of the telluric features at the airmass of observation. The integrated water vapour (IWV) was 1.78~mm and 1.01~mm for the data taken on the nights of  July 30 and August 16, 2017, respectively. It it unknown for the earlier observations. We find that the contamination from atmospheric water is $\sim$1, $\sim$1, $\sim$50 and $\sim$3 m\AA{} in equivalent width (EW) for the redshifted NaD (5889~\AA), NaD (5895~\AA), DIB (5780~\AA), and DIB (6284~\AA) lines at an airmass of 1.15 (which corresponds to that of all the observations plus or minus 0.05), and a IWV of 2.5~mm which is the median for Paranal. The integration limits were $\pm$260 km~s$^{-1}$ in the redshifted frame.  These values are lower than the threshold below which the aforementioned features are rejected, as described in Sect. \ref{sec:Absorption}.

The science exposures were combined to include the data from all 24
integral field units, corrected for atmospheric refraction, and flux
calibrated.  For the exposures with the corresponding offset sky fields, the sky
lines were initially fitted using those data (\texttt{muse\_create\_sky}) and a sky
continuum was produced as well, using the 2--87 percentile of the sky field.
The final sky-line fit was then carried out in the science exposure, using the
2--17th percentile data. The continuum was simply subtracted. For the fields
without separate sky the 2--17th percentile of the image was used to both
construct the sky continuum as well as carrying out the final sky-line fit. The
science data were then further corrected to barycentric velocities and for
distortions using an astrometric calibration, before writing an initial cube
and white-light image.
For exposure combination we corrected offsets by generating an image with point
sources using the Gaia EDR3 astrometric catalog \citep{2021A&A...649A...2L} and
cross-correlating all detections in the MUSE white-light images with the
\texttt{muse\_exp\_align} recipe. Using these offsets we were then able to
create the final datacube, combining all usable exposures, weighted by
exposure time and auto-guider seeing, with a wavelength range of
4750--9300\,\AA. This cube was again sampled like the raw data. We estimated the
absolute astrometric accuracy to be about $\pm0\farcs1$ over the full field, by
overplotting the Gaia catalog over the area again. The cube does, however, not
form a fully contiguous field, since the observed pointings did not form a
rectangular area on the sky and left gaps (see Fig. \ref{fig:datacube}).
The cube also has different depths at different positions and the image quality
varies slightly across the field (Moffat FWHM $0\farcs86\pm0\farcs10$ corresponding to an on-sky resolution of 434$\pm$50 pc at the distance of the AGNs).
In this research, we focused on the central section of \object{NGC 6240}. This is a portion of the data cube sampling an area of $\sim 32.2~ \mathrm{kpc} \times 32.2~ \mathrm{kpc}$. This zoomed in section can be seen in Fig. \ref{fig:zoomcube}, and is indicated with a green square in Fig. \ref{fig:datacube}.
Finally, before any further analysis, the final cube was corrected for a Galactic extinction $A_V=0.209$ \citep{Schlafly2011}, assuming a $R_V=3.1$ and the \citet{Cardelli1989} extinction law.

\section{Processing the data}\label{sec:method}

\subsection{Tessellation of the data}

\begin{figure}
    \centering
    \resizebox{\hsize}{!}{\includegraphics{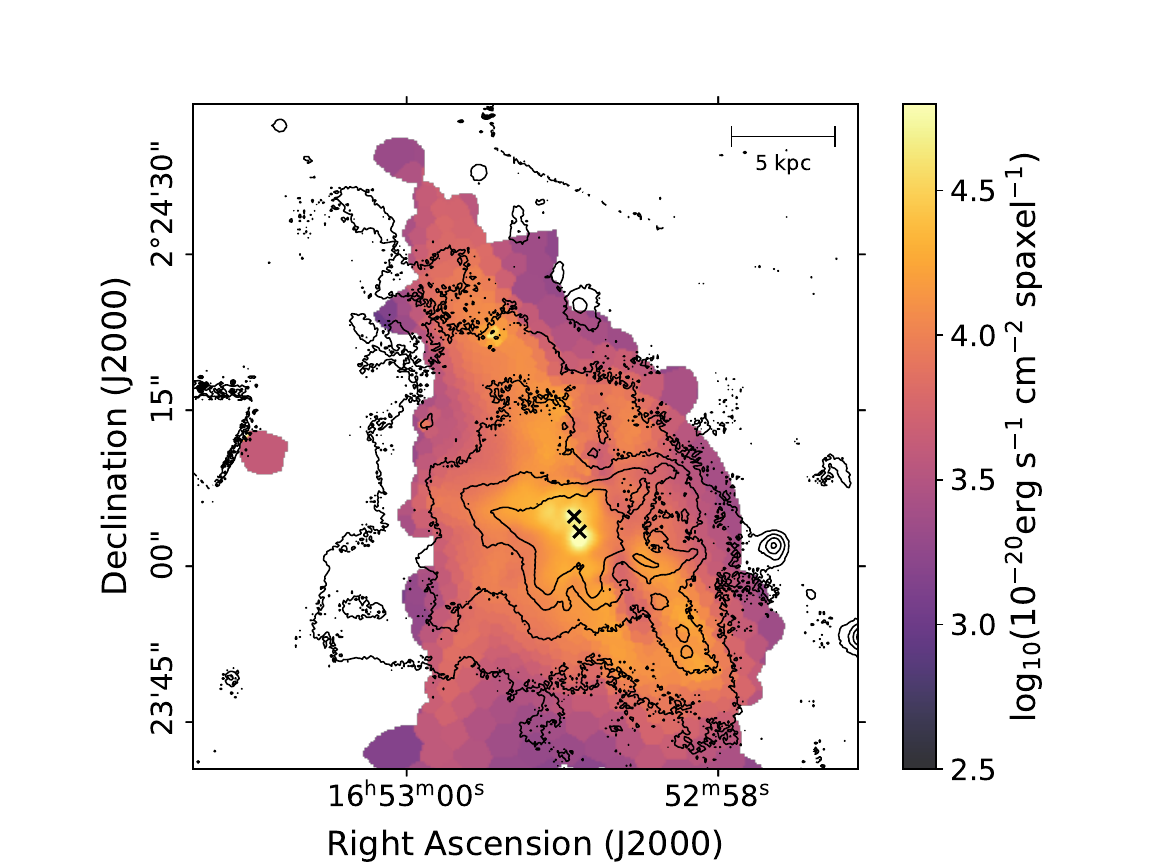}}
    \caption{Flux map of our tesselated zoomed cube calculated in a section of line free emission with range $\lambda 5700~ \AA- \lambda5850~ \AA$. The black contours trace the H$\alpha$ emission and the black crosses show the locations of the AGNs \citep{Komossa2003}.}
    \label{fig:zoomcube}
\end{figure}

In order to increase the S/N we tessellated our data, coadding the spectra, using the Python module \texttt{Vorbin V3.1.5} \citep{Cappellari2003}\footnote{\url{https://pypi.org/project/vorbin/}}. For calculations on the S/N we used the feature-free observed spectral range of 5623 \r{A} - 5986 \r{A} and a target S/N of 100, since DIBs are rather faint features. The result of this tessellation can be seen in Fig. \ref{fig:zoomcube}.

\begin{figure*}[th]
    \centering
    \includegraphics[width=17cm]{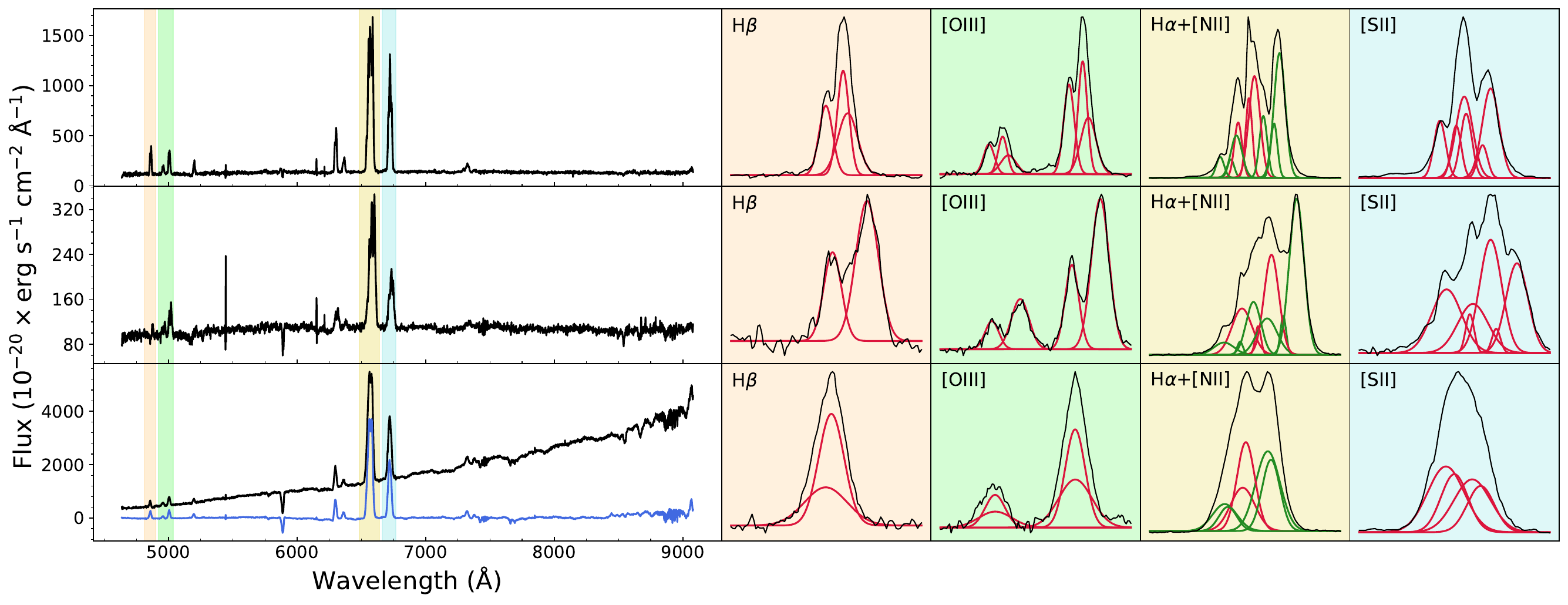}
    \caption{Sample of different \object{NGC6240} rest-frame spectra and their fits. The left hand side of the figure shows the binned and unmodeled spectra, in black. The colored regions indicate the important emission features, which are zoomed in on the right including their fits. In the \ha+\nii{} complex the green fits correspond with the \nii{} doublet. In the bottom figure, the blue spectrum represents the residual, where the stellar model is subtracted from the binned data.
    From top to bottom, these spectra are respectively measured in the regions: southeast of the AGNs, a dusty region west of the AGNs, and a bin at the location of the AGNs.}
    \label{fig:Spectrum}
\end{figure*}

\subsection{Modeling the stellar continuum \label{secstelmod}}
In a composite galaxy spectrum, there is mixed information from the stars, the interstellar medium (i.e., gas and dust) and sometimes even an AGN. To adequately derive the properties of each component, the information needs to be disentangled. In our case, we approached this by modeling the stellar continuum using the Python module \texttt{pPXF V8.0.2} described in \citet{Cappellari2003} and upgraded in \citet{Cappellari2017}\footnote{\url{https://pypi.org/project/ppxf/}}. This module fits the stellar continuum by a convolution of templates of stellar populations and a line of sight velocity distribution. For the stellar population templates, we used the {\sc E-MILES} library included in \texttt{pPXF} \citep{Vazdekis2016}. In our model there are a total of six different metallicities (-1.71, -1.31, -0.71, -0.4, 0.0, and +0.22 [M/H]) and a total of 25 different ages above 30 Myr. This results in a total of 150 stellar population templates.
As a LIRG, \object{NGC 6240} has large amounts of dust and, as a result, the attenuation is expected to be far from negligible. Thus, we added a reddening component assuming a \citet{Calzetti2000} reddening law with $R_V$=3.1.
Then, the modeled continuum model is either subtracted from our data for emission line fitting purposes (resulting in spectra with only the non-stellar residuals; see Sect. \ref{sec:Emission}) or divided for absorption features fitting (resulting in our data being normalized to one; see Sect. \ref{sec:Absorption}). An example of an emission-line residual is shown in the bottom left panel of Fig. \ref{fig:Spectrum} in blue.

\subsection{Derivation of spectral features}
We are not only interested in detecting and mapping DIBs, but also exploring their relations with other components and properties of the ISM. Thus, several quantities need to be measured from the spectra. Specifically, these are i) the equivalent width (EW) of three strong DIBs (i.e., $\lambda$5780, $\lambda$5797 and $\lambda$6284); ii) the EW of the sodium doublet (\ion{Na}{i} D) at $\lambda$5890\r{A}, $\lambda$5896\r{A}, to be used as a tracer of neutral gas; iii) the flux of the hydrogen recombination lines \ha{} and \hb{} to map the gas reddening in the system; and iv) some strong collisionally excited lines, such as \oiii $\lambda\lambda$4959,5007 and \sii $\lambda\lambda$6716,6731 used to constrain the fit of the hydrogen recombination lines.

In all cases, we modeled the spectral features using one or more Gaussian functions with help of the Python module \texttt{LMFIT} \citep{Newville2014}\footnote{\url{https://lmfit.github.io/lmfit-py/}}. However, these spectral features all present a diversity of characteristics (i.e., both isolated and blended, both in emission and in absorption, and both strong and faint features; see Fig.~\ref{fig:Spectrum} and Fig.~\ref{fig:DIBFITS}) and, thus, a tailored strategy had to be followed depending on the spectral feature of interest. In the next two sections, we describe our strategy for the features in emission (Sect. \ref{sec:Emission}) and in absorption (Sect. \ref{sec:Absorption}).

\subsubsection{Emission line fitting}\label{sec:Emission}

\object{NGC 6240} is an advanced galaxy merger that contains two AGNs, as well as a large amount of star formation. This implies a large variety of spectra, both in terms of line profiles and in terms of line ratios over the mapped area.
The \ha+\nii$\lambda\lambda$6548,6584 emission lines are often blended together due to their velocity width, which makes disentangling the flux solely emitted by H$\alpha$ challenging. Simple strategies (i.e., simulation of narrow filter images or modeling of the emission lines by a single Gaussian) cannot be applicable all over the system and a more sophisticated strategy is therefore required. 

We describe our procedure in the following. We fit less blended (but still strong) emission features aiming at using their derived parameters as constraints and initial guesses on the \ha+\nii{} complex. Specifically, we used the \oiii$\lambda\lambda$4959,5007 (preferentially) and \sii$\lambda\lambda$6716,6731 doublets. Our procedure assumes that these features lie in the same part of the ISM as \ha+\nii$\lambda\lambda$6548,6584 emission lines.
First, we start by fitting the \oiii{} doublet by a single Gaussian profile using sensible initial conditions for the flux, central wavelength, and width ($\sigma= 2$\AA). Thereafter, we allowed for up to three independent Gaussian profiles to be added.
The components were added sequentially and at each step, the quality of the fit with $N$ and $N+1$ components were compared using the Akaike information criterion \citep[$\Delta$AIC,][]{Akaike1974}, following a scheme similar to the one presented by \citet{Bosch2019}. We considered that using $N+1$ components significantly improved the fit (and thus selected that as the best option) when $\Delta$AIC$>$10.
We allowed for up to three components, since this was enough to reproduce the line profile in most of the spaxels reasonably well and to recover a reasonable measurement of the total line flux.

Then, we turned to the \sii{} doublet and proceed in a similar manner, as  \oiii{} is  poorly defined in some cases and is therefore unable to provide adequate constraints on the H$\alpha$+\nii{}  complex. The \sii{} doublet is more visible in most areas of the galaxy, but it is usually more challenging to fit because the two emission lines are closer in wavelength.

In those cases where both the \oiii{} and \sii{} doublets, could be reasonably fitted, we used again the AIC criterion to decide which parameters should be used as constraints for the H$\alpha$+\nii{} complex.
In a few cases, neither the \oiii{} doublet nor the \sii{} doublet could reasonably be fitted, since none of these emission lines were particularly strong. This happened chiefly in the outermost parts of the system. However, in these regions, the line profiles of the H$\alpha$+\nii{} complex were relatively simple and could be easily fitted without the additional input from the other lines. 

A representative selection of fits are presented on the right of Fig. \ref{fig:Spectrum}. This figure show the complexity present in some tiles. Especially the middle row shows the necessity of fitting multiple Gaussian profiles, as there are two clearly distinguishable components. In all three examples the \ha+\nii{} complex show very strong blending, which indicating the necessity to constrain the fits using the \oiii{} or \sii{} doublet.
It should be noted that we did not explicitly model the broad H$\alpha$ component in the nuclear regions, which are typical for AGNs \citep{Woo2014}. This introduces an additional uncertainty that only affects  the few spaxels in the closest vicinity of the two AGNs.
To obtain the total integrated flux, we integrated over the independent Gaussians and add them together. Errors for the emission line features and all the derived quantities hereafter were estimated via the propagation of errors, using the errors of the fitted parameters provided by \texttt{LMFIT}.

\subsubsection{Absorption features fitting}\label{sec:Absorption}

\begin{figure*}[t]
    \centering
   \includegraphics[width=17cm]{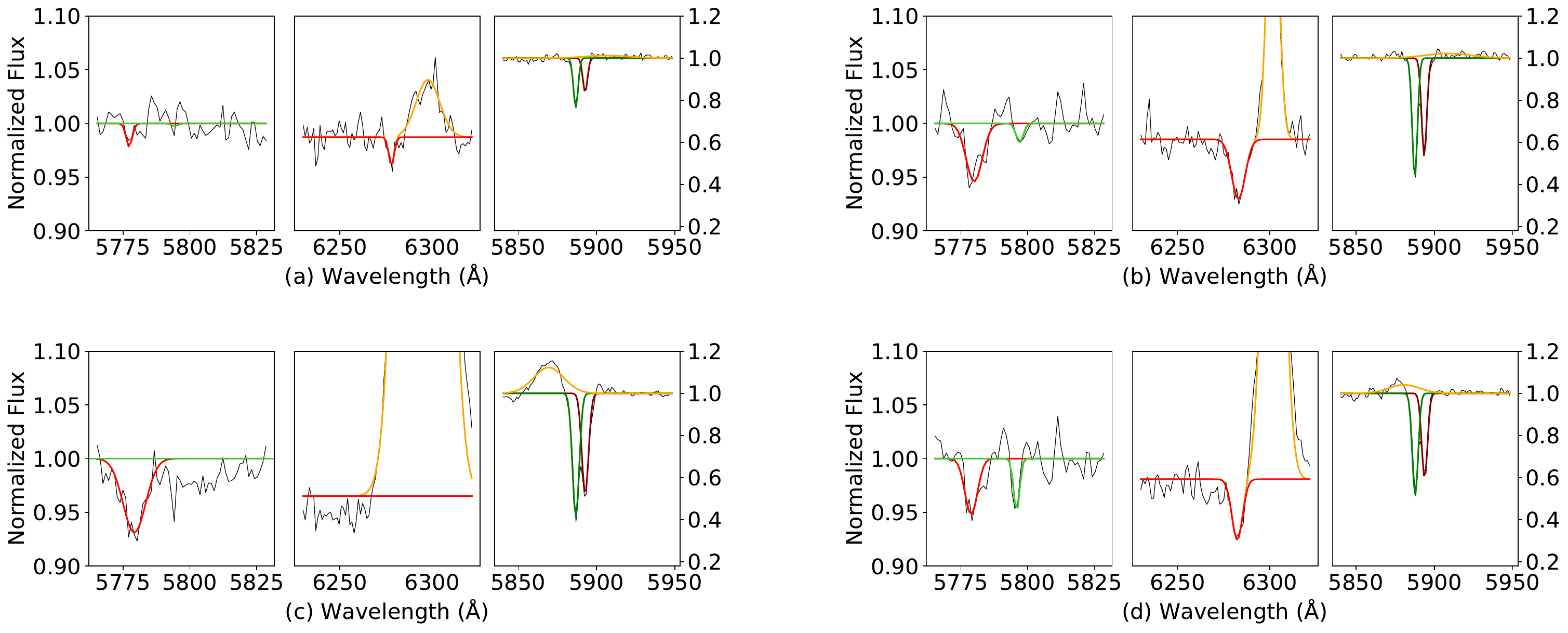}
    
    \caption{Collection of fits to the DIBs and \ion{Na}{i} D doublet in \object{NGC6240} rest-frame in four different locations. In all four figures, the left panel presents the DIBs $\lambda$5780 (red) and $\lambda$5797 (green),  the middle one shows the DIB$\lambda$6284, and the right plot displays the \ion{Na}{i} D doublet. Emission lines are presented in orange. The two panels containing the DIBs share the same y-axis. 
    } 
    \label{fig:DIBFITS}
\end{figure*}

For the fitting of the absorption features (i.e., the DIBs and the Na{}I D doublet), we used the wavelength and line width of \ha{} to provide an initial guess for the fit.
Besides, for the DIBs, we imposed them to be always in absorption and that EW($\lambda$5797)$<$EW($\lambda$5780) \citep{Kos2013,Cox2017}.
As for the sodium doublet, we used as initial condition  EW(D$_2$) / EW(D$_1$) = 1.98, the expected value for the optically thick limit \citep[e.g.,][]{Puspitarini2014}\footnote{ \url{https://physics.nist.gov/PhysRefData/Handbook/Tables/sodiumtable2.htm}}.
However, we allowed this ratio to vary, since given the amount of material in the ISM these lines may be prone to saturation in several areas of the system \citep{Poznanski2012}. 
In contrast to the fitting of emission lines, we only used a single Gaussian to fit a given absorption feature.
In the case of the DIBs, they are extremely faint features and even at high S/N of the binned spectra, it is not possible to disentangle several components. Admittedly, a handful of DIBs (not including the $\lambda$6284 DIB) are known to have more complex substructures \citep{Kos2017, MacIsaac2022} and one may argue for using a different functional form. However, at the MUSE spectral resolution, it is not possible to resolve the band profile and the single Gaussian was sufficient to recover the DIB equivalent width.
As for the sodium doublet, this is a much stronger spectral feature and an inspection of the data suggests that a detailed kinematic analysis could benefit from using multiple Gaussians to reproduce each line profile. However, for the purposes of this work (i.e., estimating its equivalent width and comparing it with that of the DIBs), one Gaussian per line suffices.
In addition, the fitting also includes modeling of nearby emission lines. Specifically, these were the \hei$\lambda5786$ emission line for the sodium doublet and the \oi$\lambda6300$ emission line for the DIB $\lambda6284$.

We found that the sodium doublet presented a P-Cygni profile with components in both absorption (blue) and emission (redder) at some locations in our covered area. The sodium doublet with P-Cygni profile has previously been detected in a few other systems \citep[e.g.,][]{Perna2019} and has been typically attributed to outflows \citep[e.g.,][]{Prochaska2011}. In our particular case, this happened mostly at the locations where we did not measure the DIBs. Since our aim was to compare the sodium and DIB emission only in regions where both are detected, we did not make any attempt to model this P-Cygni profile. 

Since absorption features in general (and DIBs in particular) are faint, extra care should be taken to guarantee a true detection. 
In particular, we were conservative and rejected any measurement with EW<70~m\AA,{} which we estimated as the lowest threshold for having a reliable measurement.
For the \ion{Na}{i} D features we ignored any fit with an EW<100 m\r{A}.
Besides, we removed those cases with abnormally high errors on the parameters of the Gaussian.
Finally, we inspected all remaining spectra visually and manually removed any unreliable fits.
Figure \ref{fig:DIBFITS} presents some examples of fits for both the DIBs and the \ion{Na}{i} D doublet. The two DIBs $\lambda5780,\lambda5797$, the DIB $\lambda6284$ and the \ion{Na}{i} D doublet are shown in the ranges of $5765 \r{A}-5828\r{A}$, $6230 \r{A}-6322\r{A}$, and $5840 \r{A}-5948\r{A}$, for four different environments. The top-left example is from an area where DIB detections were discarded given their low EW. The other three figures present detection of at least one DIB: The bottom-left figure shows a detection of DIB$\lambda$5780, while showing an example of unmeasurable DIB$\lambda$6284 due to a highly broadened \oi-line, as well as the \hei{} line strongly blended with the \ion{Na}{i} D doublet. The top-right figure shows a detection of DIBs $\lambda5780$ and $\lambda$6284, but not $\lambda$5797 (as it is not strong enough). The bottom-right example is a spectrum with detection in all the absorption features under consideration.

We estimated the errors of the absorption features using Eq. 6 of \citet[]{Cayrel1988}:

\begin{equation}
    \delta \mathrm{EW} = 1.5 \frac{\sqrt{\Delta \lambda \cdot \mathrm{FWHM}}}{S/N}
    \label{eq:DIBerr}
,\end{equation}
where $\Delta \lambda$ is the constant spectral pixel size, with 1.25 \r{A} for MUSE.  The full width at half maximum (FWHM) is given in pixels.
This results in uncertainties ranging from $\sim 20$~m\AA{} to $\sim 50$~m\AA.

\section{Results}\label{sec:results}
\subsection{DIB maps}\label{sec:Dibmaps}

\begin{figure}[h]
    \centering
    \resizebox{\hsize}{!}{\includegraphics{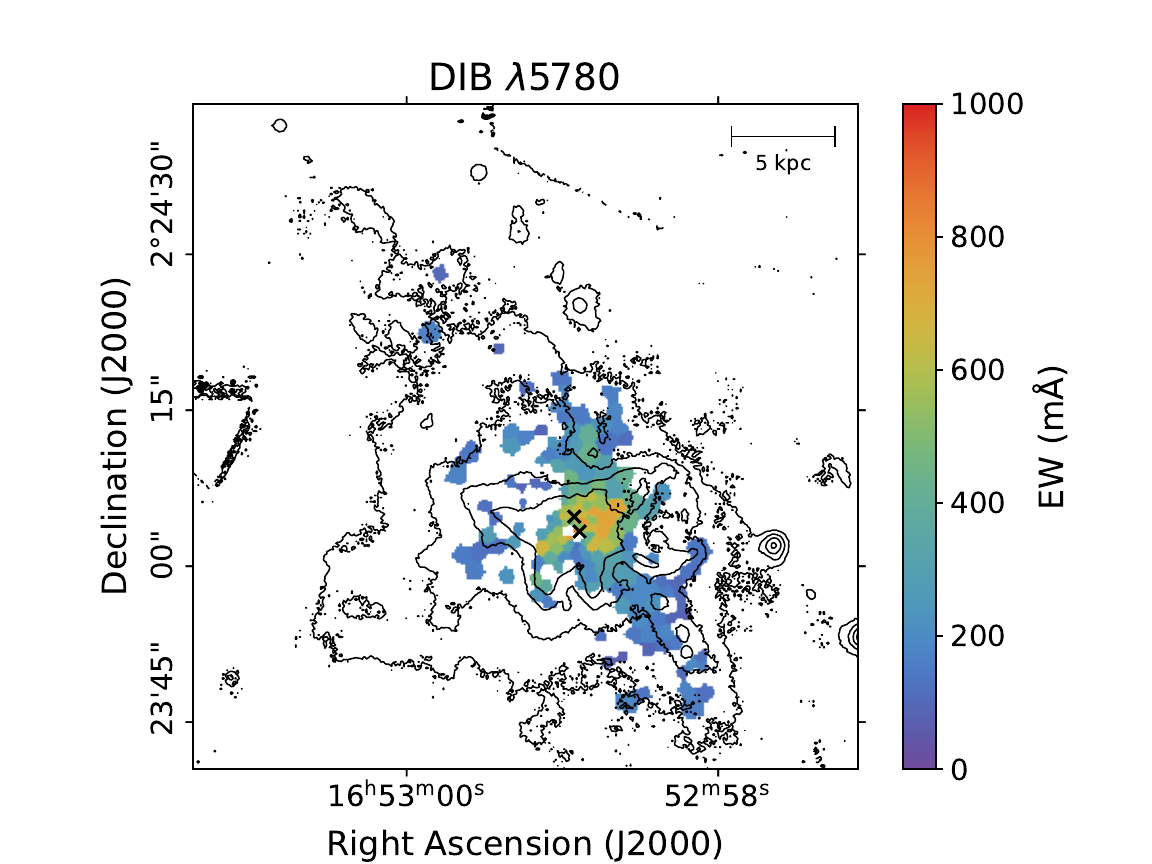}}\\
    \resizebox{\hsize}{!}{\includegraphics{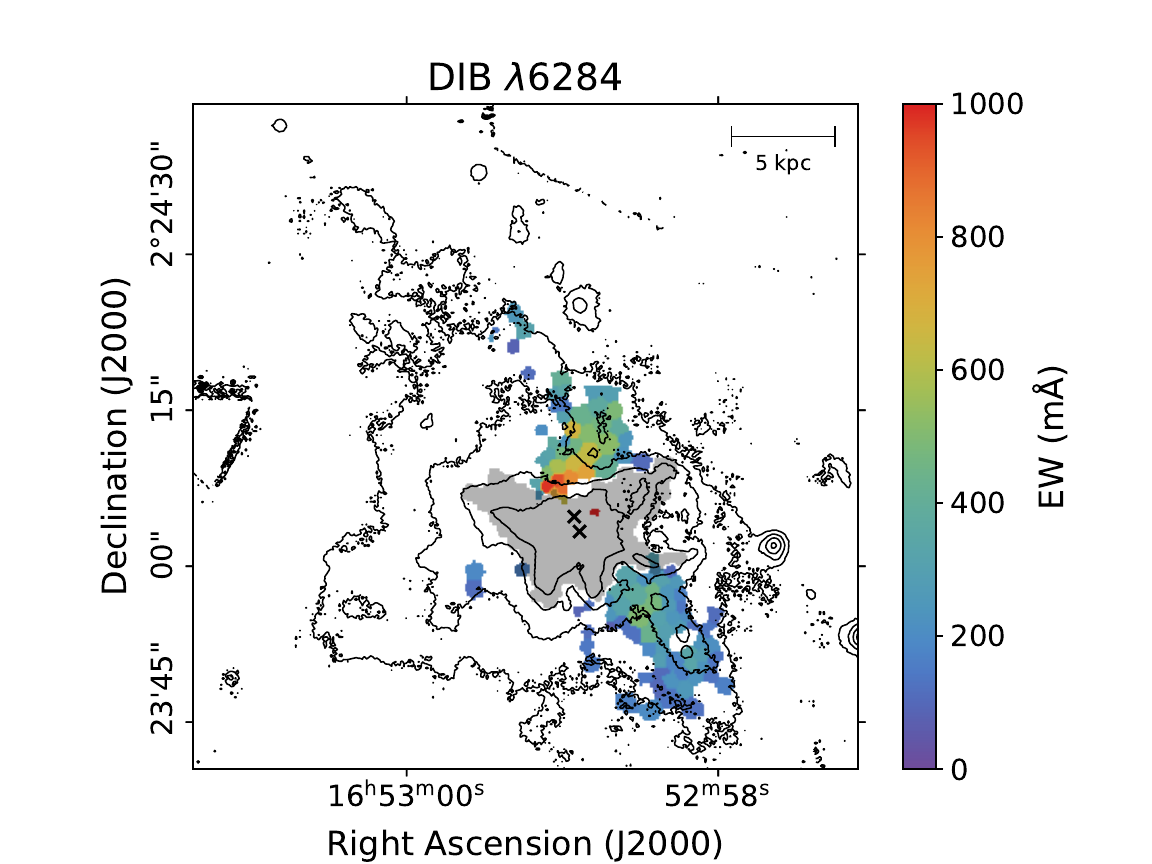}}
    \caption{Maps of detections of DIB $\lambda 5780$ (top) and DIB$\lambda$6284 (bottom). In the bottom panel, the gray area is the region where EW(\oi$\lambda6300$)$\geq 10$\AA. The lack of DIB$\lambda$6284 detections in these tiles is likely an observational artefact, as the line is blended with the broader \oi$\lambda6300$ feature. The black contours trace the H$\alpha$ emission and the black crosses show the locations of the AGNs \citep{Komossa2003}.}
    \label{fig:DIBmaps}
\end{figure}

We measured the DIBs at $\lambda$5780 and $\lambda$6284 in more than 100 independent lines of sight. 
A summary of the measurements is presented in Table \ref{table:DIBDet}.
The maps with the equivalent width derived from these measurements are presented in Fig. \ref{fig:DIBmaps}.
The DIB at $\lambda$5797 was only marginally detected in 13 tiles and will therefore not be discussed hereafter. However, it is worth mentioning that this marginal detection calls for further and deeper observations aiming at deriving maps similar to those presented in Fig. \ref{fig:DIBmaps}, allowing us to explore the relation between this and the two DIBs studied here. This is particularly interesting since the ratio $\lambda$5780/$\lambda$5797 has been proposed to trace the level of ionization
(\citealp[e.g.,][]{Krelowski1999, Sollerman2005, Vos2011, Ensor2017, Elyajouri2017}, although see \citealp{Lai2020} for a rebuttal), which could turn out to be an engaging study, given a mergers' active star-formation and complex kinematics.

\begin{table}[t]              
\caption{\label{table:DIBDet}Basic data on the DIB$\lambda$5780 and DIB$\lambda$6284 detections.}
\centering                                      
\begin{tabular}{ccccc}          
\hline\hline                        
\smallskip
DIB & $N_\mathrm{tiles}$ & $A_\mathrm{Tot}$ & $\overline{n}$/tile & $\overline{A}$/tile \\    
 & & ($\mathrm{kpc}^2$) & & ($\mathrm{kpc}^2$) \\
\hline                                   
    $\lambda$5780 & 171 & 76.63 &40 & 0.41 \\      
    $\lambda$6284 & 109 & 57.71 &48 & 0.49 \\
\hline                                             
\end{tabular}

\tablefoot{Column 2 contains the total amount of tiles where these DIBs were detected. The total area where they were mapped is presented in column 3. The median number of spaxels per tile is listed in column 4. The median physical size of a tile is presented in column 5.}  
\end{table}

The DIB $\lambda$5780 is detected in a contiguous elongated region close to the AGNs, stretching over 15 kpc from the north to the south-west, covering a total area of 76.63 kpc$^2$ (upper panel in Fig. \ref{fig:DIBmaps}).
The EW($\lambda$5780) is the highest at the center of the system. However, the peak ($\sim 800$ m\r{A}) presents an offset of 2 kpc toward the east with respect to the two nuclei. From there on, the strength of the DIB decreases gradually, going outwards.
In addition, there are some detections to the west of the two AGNs but they are not in a large continuous field and relatively scant. Because of their isolation and low equivalent width, these should be taken with a grain of salt.
The areas where the DIB at $\lambda$6284 was detected is presented in the lower panel of Fig. \ref{fig:DIBmaps}.
In contrast to the DIB $\lambda$5780, this DIB is detected in two separate swathes to the north and the south-west, with a total area covering 59.78 kpc$^{2}$.
The southern region is the largest, and has an area of $\sim$31.41 kpc$^2$. The northern region is smaller and has an area of $\sim$21.22 kpc$^2$. However the equivalent width there is stronger reaching values as high as $\sim1000$~m\r{A}.
The lack of detections in the vicinity of the AGNs is an observational effect, and does not necessarily imply the absence of the carrier causing this DIB.
This is because this DIB is very close in wavelength to the \oi$\lambda$6300 emission line,
which becomes very broad in the surroundings of the AGN pair (see the gray area in Fig. \ref{fig:DIBmaps}). As a consequence, if the extremely faint DIB (EW typically measured in m\AA) existed, it would become completely buried in the wings of the much stronger \oi{} line (EW can range from 1 to 100 $\mathrm{\AA}$) and thus, impossible to be disentangled from it (see bottom left figure of Fig. \ref{fig:DIBFITS}). This broadening of spectral features happens in regions where the gas is very turbulent or very hot, which is the case for this region in \object{NGC 6240} \citep{Muller2018, Yoshida2016}.
Even if  the DIB$\lambda$5780 \r{A} was previously detected by \citep{Heckman2000} in an integrated spectrum for the system, this is the first time that the spatial distribution of this DIB is measured in the system. In their study, \citet{Heckman2000} obtained an equivalent width of $640 \pm 70$ m$\mathrm{\AA}$. This value nicely compares with the equivalent width we map in the vicinity of the two AGN.
\citet{Heckman2000} also predicted an EW($\lambda$6284) of 1408$\pm$154 m$\mathrm{\AA}$ using the ratio of equivalent width (W($\lambda$6284)/W($\lambda$5780)=2.2) provided by \citet{Chlewicki1986, Porceddu1992}. This predicted value is larger than what we measured. The reason for this is likely due to the inability to measure DIB$\lambda$6284 in the center, where the DIB$\lambda$5780 compares nicely with \citet{Heckman2000}.
Our results confirm that DIB$\lambda$6284 is de facto in NGC 6240, even if we can only detect it outside the center. We mapped it for the first time in this system and beyond the Local Group.

Leaving aside the very center of the system, where the \oi{} line prevents us from measuring the DIB$\lambda$6284, we find that both $\lambda$5780 and $\lambda$6284 DIBs display a similar structure. They are detected in the same regions to the north and south-west of the AGNs and are both also strongest when going toward the inner parts of the system. DIB$\lambda$6284 appears, when detected, to be the strongest of the two DIBs. This is line with previous research \citep[e.g.,][]{Friedman2011, Raimond2012}, in which this DIB was also found to be stronger.

\subsection{Reddening maps}\label{sec:reddeningmaps}

There is a well established correlation between the strength of the DIBs and the interstellar extinction, in our Galaxy \citep[e.g.,][]{Capitanio2017,Elyajouri2016,Lan2015} and in other galaxies \citep[see][and references therein]{MonrealIbero2018}. Since here we are working with a different environment, it is worth exploring how the strength and distribution of the DIBs identified in \object{NGC 6240} compares with dust attenuation suffered by both the stellar component and the ionized gas, as traced by the reddening, $E(B-V)$.
Figure \ref{fig:RedMaps} contains a map for both reddening types. Both maps cover the same range in reddening to make the comparison between them easier. The maps are discussed in the following sections.
\subsubsection{Reddening derived from the stellar continuum}\label{sec:Stelred}

\begin{figure}[h]
    \centering
    \resizebox{\hsize}{!}{\includegraphics{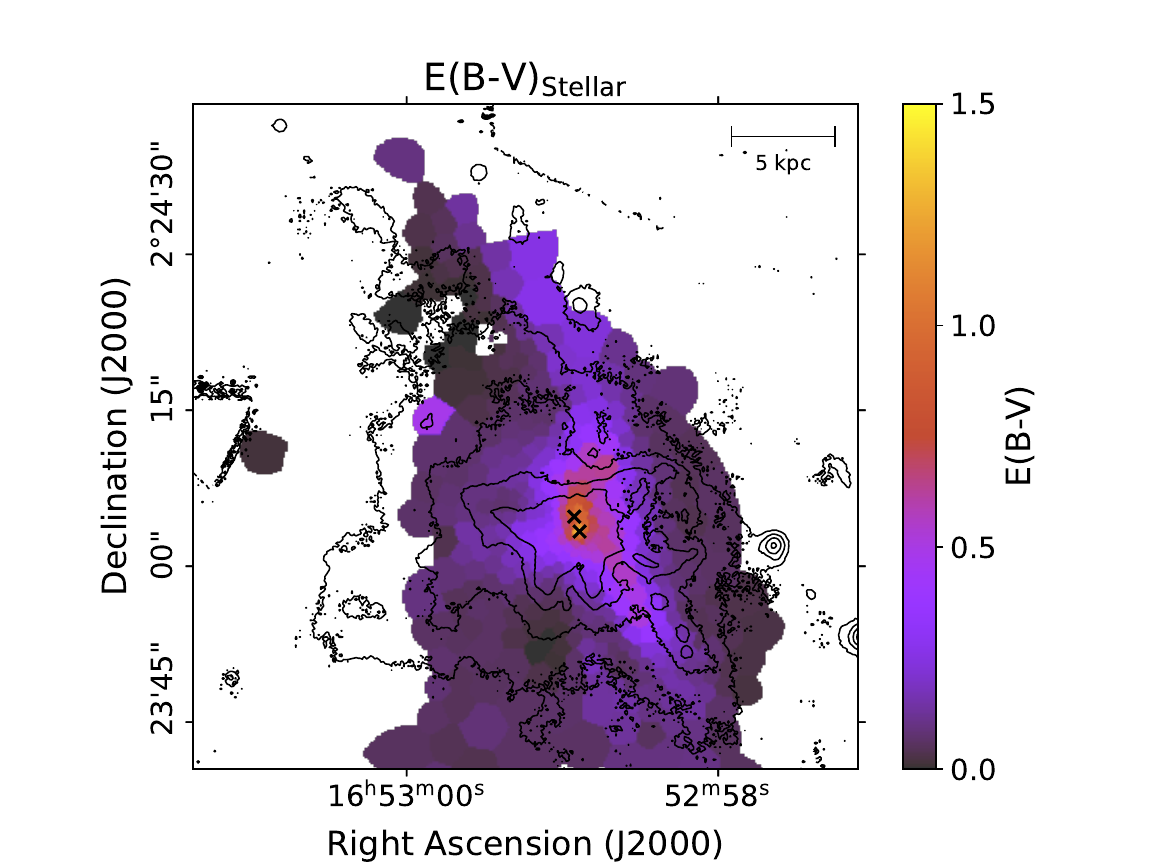}}
    \resizebox{\hsize}{!}{\includegraphics{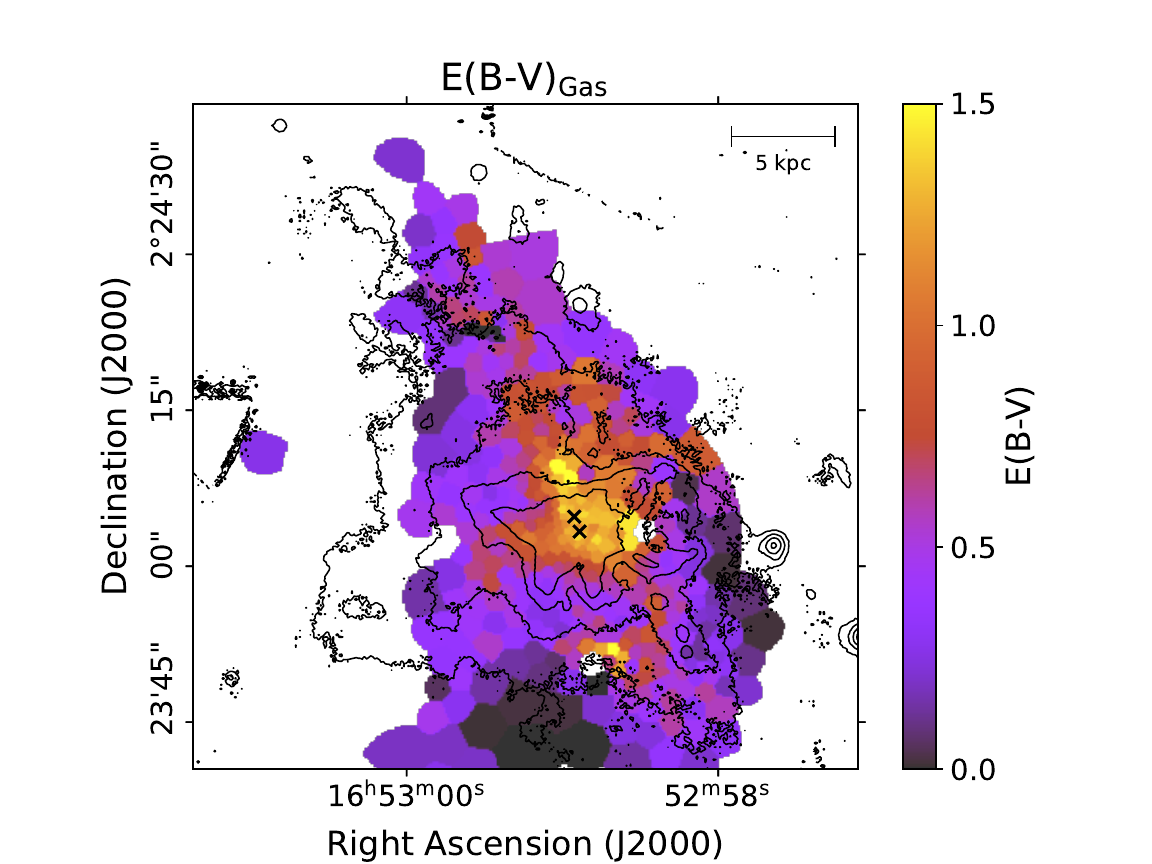}}
    \caption{Maps of the $E(B-V)$ for the stellar continuum and diffuse gas. The top figure shows the attenuation of the stellar continuum. The bottom figure shows the attenuation as derived from the Balmer decrement. The black contours trace H$\alpha$ emission. The locations of the AGNs are marked with black crosses \citep{Komossa2003}.
    }
    \label{fig:RedMaps}
\end{figure}

The reddening of the stellar continuum was obtained as a by-product of the modeling of the stellar continuum with \texttt{pPXF} \citep[][see Sect. \ref{secstelmod} for details]{Cappellari2017}, 
and is presented in the top panel of Fig. \ref{fig:RedMaps}.
Reddening is the highest in the central area, near the pair of AGNs, where it reaches values of $E(B-V)\sim$1.0. These are comparable to previous reddening determinations in this area \citep[][]{Max2005,Scoville2015, Muller2018}.
When moving outwards, reddening consistently decreases, although not by the same amount in every direction.
A $\sim$25 kpc-length lane of dust, seen as a location with relatively high reddening, crosses the mapped area from north to south. 
Both results, namely, the highest attenuation in the center of the system and dust lanes is similar to results found in other dusty advanced mergers such as \object{Arp 220} \citep{Perna2020}.

\subsubsection{Reddening derived from the ionized gas}\label{sec:gasred}
We also derived the reddening using the Balmer decrement, assuming case B recombination, with a temperature of $T_\mathrm{e}=10^{4}\ \mathrm{K}$ and electron density of $n_\mathrm{e} = 10^2\ \mathrm{cm}^{-3}$ \citep{Osterbrockbook}. 
This can be expressed as the following equation:
\begin{equation}
    \label{eq:gasreddening}
    \centering
    E(B-V)_\mathrm{Gas} = \frac{2.5}{\mathrm{k}(\mathrm{H}\beta)-\mathrm{k}(\mathrm{H}\alpha)} \cdot \log_{10}\left(\frac{\mathrm{H}\alpha / \mathrm{H}\beta}{2.86}\right)
,\end{equation}
where $E(B-V)_\mathrm{Gas}$ is the color excess value for the ionized gas. We assumed a \citet{Calzetti2000} reddening law for the wavelength dependent extinction parameter k($\lambda$). The corresponding map is presented in the bottom panel of Fig. \ref{fig:RedMaps}.
Several pieces of knowledge can be inferred from this map and its comparison with the map of the stellar reddening.
In general, the reddening derived from the ionized gas is clearly higher than for the stellar continuum.
We further discuss this difference in Sect. \ref{sec:relred}.
Moreover, the overall general structure of the gas and stellar reddening are relatively similar, even if the lane of attenuation is less clearly defined for the gas reddening. Some more remarkable quantifiable differences are clearly noticeable.
Firstly, as with the stellar reddening, the gas reddening is centered and high in the vicinity of the two AGNs, reaching values of $E(B-V)\sim$1.3. However, this is not the location with the highest reddening. There are several local maxima with higher reddening spread all over the mapped area. 
The location with the highest reddening is at $\sim$2.5~kpc toward the north-northeast of the pair of AGNs (interestingly, the stellar reddening is strongest in the area a few kpc to the west of the AGNs).
Secondly, there is an area ranging from $\sim$1 to $\sim$3~kpc west of the AGNs with $E(B-V)\sim1.3$. 
Thirdly, at $\sim$7~ kpc toward the south of the southern AGN there is a third local maximum in gas reddening with $E(B-V)\sim1.5$. 
Lastly, we note relatively low values of reddening north-east of the AGN, with $E(B-V)\sim0.5$. Interestingly enough, a very strong ionized outflow is able to ionize the hydrogen and remove dust has been reported at this location \citep{Muller2018}.
The reddening in \object{NGC 6240} has already been mapped to some extent in the past.
For instance, at the smallest scales, \citet{Scoville2000} used NICMOS images and assumptions about the age of the stellar population in the system to map the reddening in a $\sim$2.4~kpc$\times$2.4~kpc area
in the closest vicinity of the two AGNs, the circumnuclear regions obscured behind high dust column densities.
Also, \citet{Medling2021} present a \ha/\hb{} line ratio (i.e., same tracer as here) map over the central $\sim$8~kpc$\times$4~kpc.
Our $E(B-V)_{\mathrm{Gas}}$ values are in good agreement with their line ratios, and extend the reddening mapping over a much larger area, of $\sim$13~kpc$\times$25~kpc. This is, to our knowledge, the most comprehensive reddening map in \object{NGC 6240}, presented so far.

\begin{figure}[h]
    \centering
    \resizebox{\hsize}{!}{\includegraphics{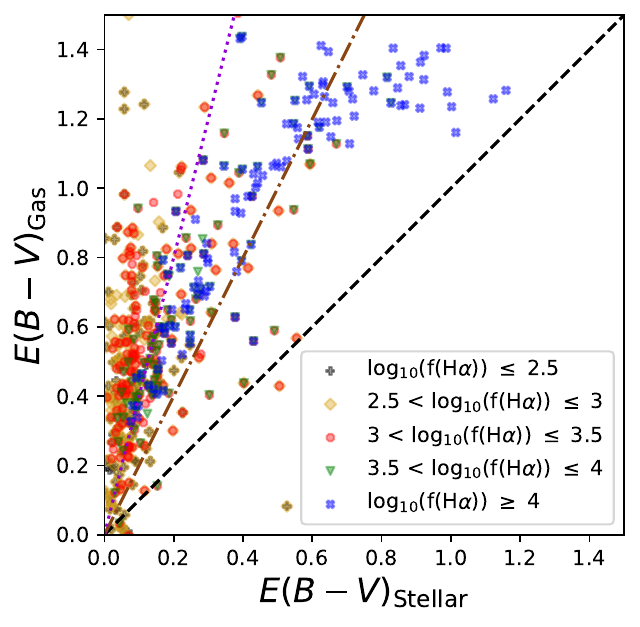}}
    \resizebox{\hsize}{!}{\includegraphics{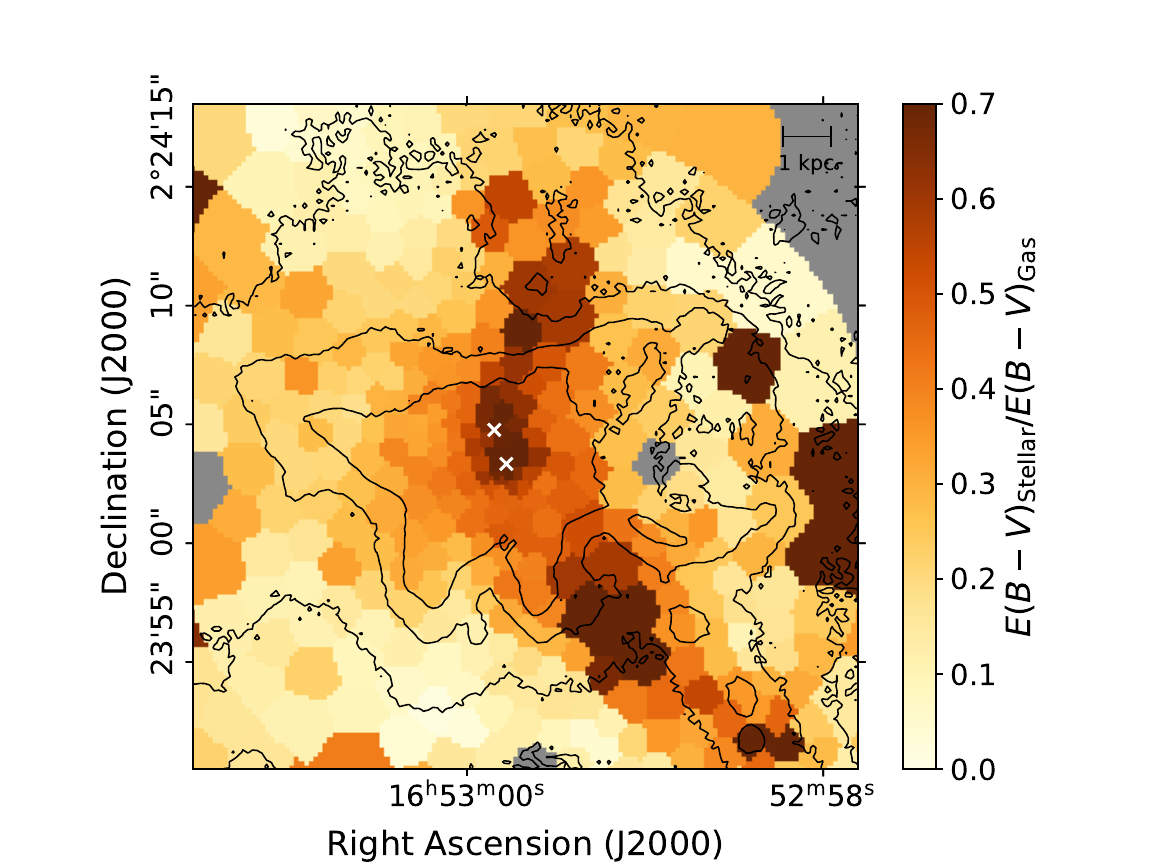}}
    \caption{Comparison of the reddening traced by the gas and the stars.
    In the plot on the top, data points are colored according to five different bins in surface brightness with units of f(H$\alpha$) $10^{-20}$ erg s$^{-1}$ cm$^{-2}$ spaxel$^{-1}$, corresponding with the black contours earlier used in the paper.
    The black dashed line indicates a one-to-one relation, the brown dash-dotted line is the relation $E(B-V)_{\mathrm{Stellar}} = 0.5\times E(B-V)_{\mathrm{Gas}}$, and the purple dotted line is the relation  $E(B-V)_{\mathrm{Stellar}} = 0.25\times E(B-V)_{\mathrm{Gas}}$. The bottom map shows the ratio of $E(B-V)_\mathrm{Stellar}/E(B-V)_\mathrm{Gas}$ in each tile. The locations of the AGNs are marked with white crosses \citep{Komossa2003}.}
    \label{fig:GasStarRedComp}
\end{figure}

\subsubsection{Relation between reddening derived from the stellar continuum and Balmer decrement}\label{sec:relred}

The stellar and gas reddening maps display a similar morphology (see Fig. \ref{fig:RedMaps}).
There is a lane of high $E(B-V)$ from north to south, being the highest in the nuclear region, and with low $E(B-V)$ values in the highly ionized AGN-driven north-eastern outflow \citep{Muller2018}. 
That said, there are significant differences between both maps. This is further illustrated in Fig. \ref{fig:GasStarRedComp}.
In the top panel, we show the relation between the stellar and gas reddening on a tile-per-tile basis with data points coloured according to 5 bins of H$\alpha$ surface brightness. Clearly, the bin with the highest surface brightness, tracing mainly the center of the system, has the highest amount of reddening in both metrics. Likewise, this is the bin where stellar and gas reddening present the most similar values (i.e., these data points are the closest to the one-to-one relation).
Moreover, when looking at the whole set of data points, the stellar and gas reddening seem to follow a nearly linear relationship up to $E(B-V)_{\mathrm{Gas}}\sim1.1$. Beyond that value, the relation between both quantities seems to flatten.

The bottom panel contains a 2D mapping of the ratio of $E(B-V)_{\mathrm{Stellar}}/E(B-V)_{\mathrm{Gas}}$ map in the innermost part of the system (white square in Fig. \ref{fig:datacube}). There, the area corresponding to the flattened portion of the relation is displayed with darker colours, and corresponds to a stellar reddening that is, at least, half of the gas reddening. It roughly does follow the main dust lane, although some portions of this lane still present large differences between both reddenings (e.g., toward the southwest of the southern nucleus).
Figure \ref{fig:GasStarRedComp} suggests a complex distribution of gas, dust, and stars, where their spatial distribution is mixed in a non-obvious fashion. 
In a scenario where stellar feedback manages to clear out the environment where stars were born as they age, one would expect younger stellar populations to experience a larger amount of attenuation than the older ones \citep{Calzetti1994}.
In a system as complex as \object{NGC 6240} and along a given line of sight, different combination of stars at different ages are expected, all of them contributing to the stellar continuum. Thus, the stellar reddening inferred from the modeling of this continuum can be understood as a (light-averaged) reddening for all the stars along a given line of sight.
On the other hand, the ionized gas is located preferentially in the vicinity of massive ionizing O stars or the AGNs. Thus, gas reddening would trace mostly the attenuation suffered by the young(est) stars.
Therefore, the ratio between the $E(B-V)_{\mathrm{Gas}}$ and $E(B-V)_{\mathrm{Stellar}}$ can be close to unity either because the stellar continuum is dominated by the emission of the young stars or because stars (independently of their age) are all shrouded behind a similar amount of gas and dust, with old stars somehow still embedded in their clouds.
\object{NGC 6240} contains large amounts of gas and dust and is actively forming stars at an extremely high rate \citep{Heckman1990,Scoville2000,Beswick2001,Yoshida2016,Paggi2022}. Thus, in principle, both situations are plausible and can act in a non-exclusive manner. The center of the galaxy, where most of the star formation is taking place, may be an example dominated by the first scenario \citep{Pasquali2004}, while in the dust lane, the second scenario could perhaps play a more significant role.

\subsection{\ion{Na}{i} D maps}
\begin{figure}[h]
    \centering
    \resizebox{\hsize}{!}{\includegraphics{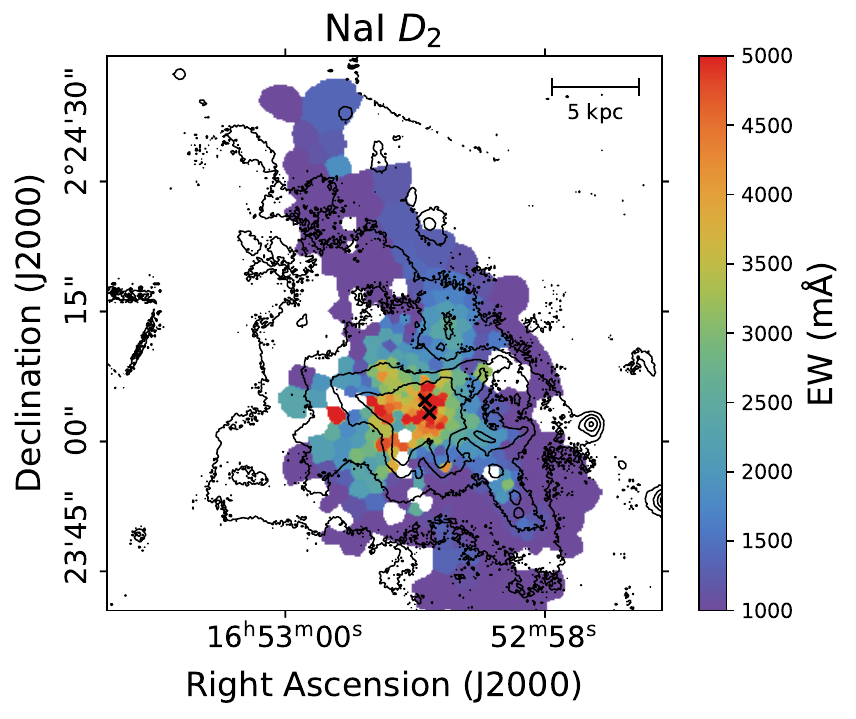}}
    \caption{Map of the NaI $\mathrm{D_2}(\lambda5890)$ line. The black contours show H$\alpha$ emission. The locations of the AGNs are marked with black crosses \citep{Komossa2003}.}
    \label{fig:NAIDmap}
\end{figure}

We  also mapped the equivalent width for the \ion{Na}{i} D absorption doublet.
Figure \ref{fig:NAIDmap} presents the map for the \ion{Na}{i} D$_2$ ($\lambda$5890) line.
The highest values ($\sim$5000 m$\AA$) are found in the central region, close to both AGNs. From this center the EW decreases when going outward in a rather steep and homogeneous manner, up to the outermost parts of the dust lane (i.e., $\sim$15~kpc to the north and $\sim$10~kpc to the southwest) where we measure values of $\sim$1000~m\AA.
Additionally, there is an area of relatively high EWs ($\sim$3000~m\AA) at $\sim$5 kpc toward the northwest of the two nuclei, roughly below the starburst-driven outflow \citep{Muller2018, Yoshida2016} and coinciding with an area of relatively high gas reddening (see Fig. \ref{fig:RedMaps}).

\section{Discussion}\label{sec:discussion}

\subsection{Relation DIBs and attenuation}\label{sec:DiscussionDIBebv}

\begin{table}[h]              
\caption{\label{table:Fits} Parameters of the fits presented in Fig.~\ref{fig:DibGascomp}, using Eq.~\ref{eq:Power}.}
\centering                                      
\begin{tabular}{cccc}          
\hline\hline                        
\smallskip
Figure & Fit & $\gamma$ & $A$ \\    
\hline                                   
   (a)&\object{NGC 6240} & $0.54 \pm 0.04$ & $0.68 \pm 0.023$ \\
   (a) &MC&$0.69 \pm 0.29$ & $0.16 \pm 0.086$\\
   (a)& No MC & $0.92 \pm 0.02$ & $0.41 \pm 0.015$ \\
   (a)& \citealp{Lan2015} & 1 $\pm$ 0.02 & 0.43 $\pm$ 0.02\\

    (b) &\object{NGC 6240} &$1 \pm 0.08$ & $0.45 \pm 0.012$\\
   (b)& MC & $0.69 \pm 0.29$ & $0.16 \pm 0.086$ \\
   (b)&No MC & $0.85 \pm 0.02$ & $0.45 \pm 0.012$ \\
   (b)& \citealp{Lan2015} & 1 $\pm$ 0.02 & 0.43 $\pm$ 0.02\\

    (c) &\object{NGC 6240} &$0.68 \pm 0.05$ & $0.86 \pm 0.087$\\
   (c)& MC & $0.59 \pm 0.27$ & $0.43 \pm 0.204$ \\
   (c)&No MC & $0.71 \pm 0.03$ & $0.86 \pm 0.047$ \\
   (c)& \citealp{Lan2015} & 0.81 $\pm$ 0.01 & 0.86 $\pm$ 0.02\\

   (d) &\object{NGC 6240} &$0.98 \pm 0.14$ & $0.41 \pm 0.031$\\
   (d)& MC & $0.59 \pm 0.27$ & $0.43 \pm 0.204$ \\
   (d)&No MC & $0.71 \pm 0.03$ & $0.95 \pm 0.047$ \\
   (d)& \citealp{Lan2015} & 0.81 $\pm$ 0.01 & 0.86 $\pm$ 0.02\\
\hline                                             
\end{tabular}
\tablefoot{Column 1 represents the subfigure and Column 2 is the specific fit. Column 3 contains the exponent and Column 4 is the constant.} 
\end{table}

\begin{figure*}[t]
    \centering
    \includegraphics[width=17cm]{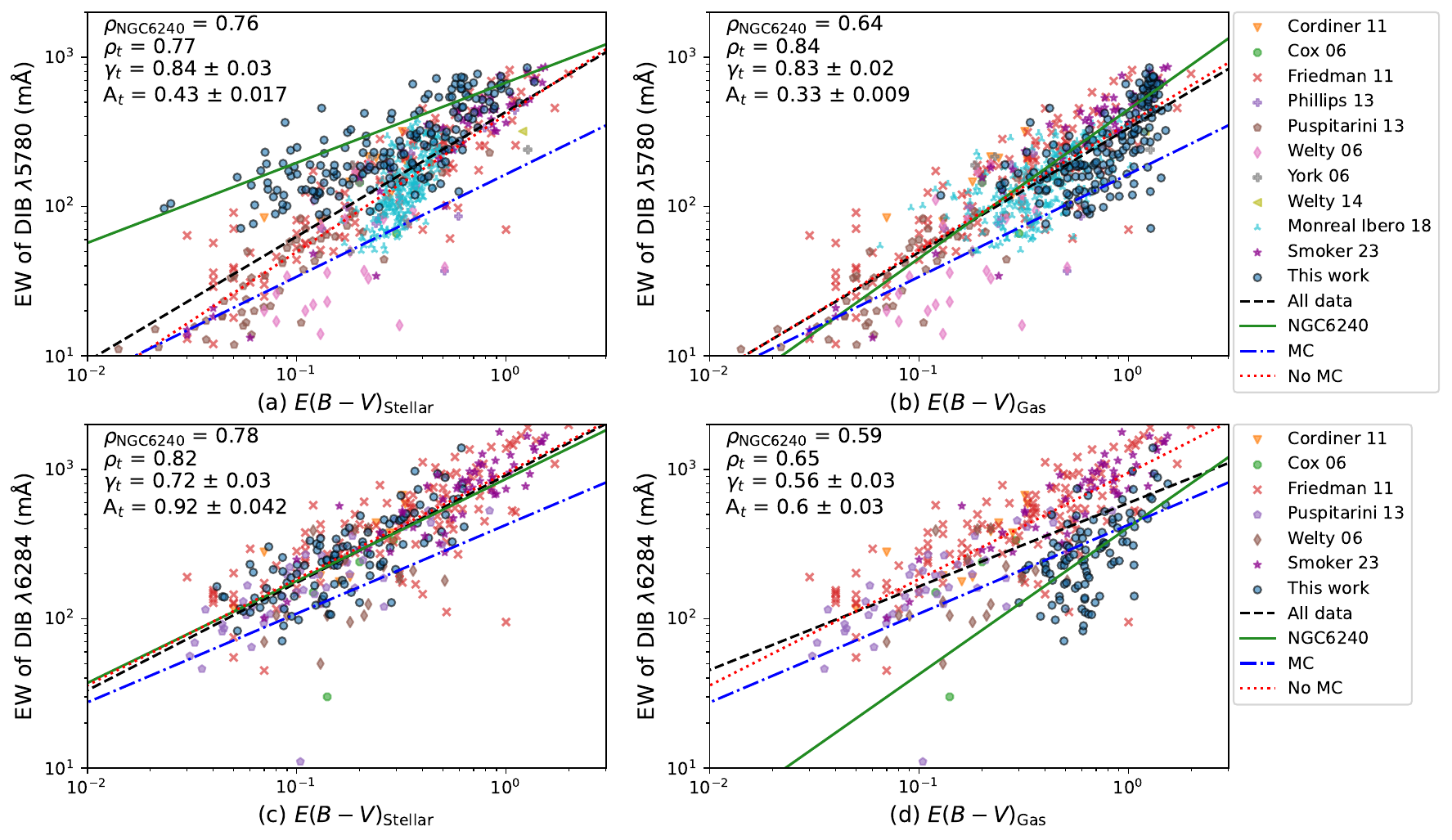}
    \caption{Comparison of the stellar and gas $E(B-V)$ and the equivalent widths of the DIBs at $\lambda$5780 and $\lambda$6284, as measured by various studies. The black (green, blue, red) dashed (solid, dash-dotted, dotted) line is the power law fitted to all available (\object{NGC 6240}, MC, previous studies except MC) data, mathematically presented in Eq. \ref{eq:Power}, using the $\gamma_t$ and A$_t$ values shown in the corresponding figure. The correlation between the EW of the DIBs and $E(B-V)$ using solely our and the total the data is also shown, by the Pearson correlation coefficient, respectively $\rho_\mathrm{\object{NGC6240}}$ and $\rho$. The top figures show the attenuation of the stellar continuum (a) and Balmer decrement (b) (Eq. \ref{eq:gasreddening}) with respect to DIB$\lambda$5780. The bottom panels show the attenuation of the stellar continuum (c) and Balmer decrement in relation with DIB$\lambda6284$. The fitting parameters for the \object{NGC6240}, MC, no MC, and the \citealp{Lan2015} fits are presented in Table \ref{table:Fits}.}
    \label{fig:DibGascomp}
\end{figure*}
In Figs. \ref{fig:DIBmaps} and \ref{fig:RedMaps}, we present maps of the DIB$\lambda$5780, DIB$\lambda$6284 and $E(B-V)$. These maps display a similar structure, where the regions with the highest DIB EWs are located in the areas with large reddening. This suggests a spatial correlation between the DIBs and reddening which is similar to detections in our Galaxy \citep[e.g.,][]{Vos2011, Baron2014, Elyajouri2016} and the Antennae Galaxy \citep{MonrealIbero2018}. 

As such, we put our results in context by comparing them with previous studies.
In particular, we use the following list of studies:
stars in the Milky Way \citep{Friedman2011, Puspitarini2013, Smoker2023}, and the Magellanic Clouds \citep{Welty2006, Cox2006}, and \object{M\,31} \citep{Cordiner2011}, supernovae \citep{Phillips2013,Welty2014}, as well as the Antennae Galaxy \citep{MonrealIbero2018}.
The compiled information is presented in Fig. \ref{fig:DibGascomp}. The two panels on the left show the relation between the DIBs and the stellar reddening, while the two panels on the right contain the comparison between the DIBs and the gas reddening.
With different nuances, depending on the DIB and whether we are using the stellar or gas reddening, our measurements in \object{NGC 6240} fall in the EW(DIB) vs. $E(B-V)$ relation delineated by previous measurements, which globally follows a linear relation (in the logarithmic stretching). This relation will be further examined by i) quantifying and comparing the strength of these relations by means of the Pearson correlation coefficient ($\rho$) and ii) fitting a power law function that relates both quantities.

We began by examining the correlation coefficients. 
In our data ($\rho_\mathrm{\object{NGC6240}}$), both DIBs have a superior correlation with the stellar reddening than with the gas reddening (\citealt{Friedman2011} propose that a minimum correlation of 0.86-0.88 is required for any physical similarity). 
This could be attributed to the broader range of $E(B-V)$ covered by the stellar reddening. Specifically, within the regions where DIBs were mapped, gas reddening spans $\sim$1~dex for both DIBs, whereas stellar reddening extends over nearly $\sim$1.5 and $\sim$2 dex for DIB$\lambda$6284 and DIB$\lambda$5780 respectively. Hence, due to the  gas reddening being measured in a smaller dynamical range it is more prone to display spurious correlations, and/or hide actual correlations. Previous work by \citet{Lan2015} finds a turnover in $\lambda$5780 and $\lambda$6284 DIB strength vs. $E(B-V)$ in Sloan survey data at $E(B-V)\sim$0.5. This is not obviously seen in our data. 
As anticipated, the correlations remain somewhat weaker compared to those obtained for the same DIBs using Galactic sightlines \citep[e.g.,][]{Kos2013, Li2019}. This underscores the difficulty in detecting DIBs beyond the Local Group.
Even if the high efficiency of the MUSE+VLT combination enabled us to map the DIB$\lambda$5780 and DIB$\lambda$6284, it came at the expense of sacrificing the typically superior spectral and spatial resolution commonly employed in Galactic studies.
Regarding the spatial resolution, this implies that small scale structure in both the DIBs could potentially be diluted during the tessellation process. Similarly, our sightlines are unlikely to be dominated by a single cloud and thus they likely span a range in physical conditions. 
Regarding the spectral resolution, residuals in the modeling of the underlying stellar continuum and, in the case of DIB$\lambda$5780, a potential contamination of the much fainter DIB$\lambda$5779 \citep[blended at this resolution;][]{Jenniskens1994, Galazutdinov2020} may be present.
Both effects may contribute to decreasing the observed intrinsic correlation.
Overall, the correlation coefficients found here compare well with those determined in the Antennae Galaxy, observed with the same instrument \citep{MonrealIbero2018}.
Figure \ref{fig:DibGascomp} also contains the correlation coefficients calculated for all the data ($\rho_t$).
In general, the degree of correlation remains roughly the same or, as is the case for the DIB$5780$-$E(B-V)_{\mathrm{Gas}}$ relation (panel b), gets even stronger, despite the large variety of sampled environments (i.e., gas metallicities, star formation rates, galaxy Hubble type, etc) and observational techniques (i.e., single line of sight vs.\ information integrated over a large area). However, the high Pearson correlation coefficient for the DIB$\lambda5780$-$E(B-V)_{\mathrm{Gas}}$ relation needs to be taken with caution. On the one hand, as mentioned in Sect.~\ref{sec:relred}, gas reddening preferentially traces the attenuation toward the locations with heavy star formation, while the stellar reddening is a better tracer of the overall extinction along a given line of sight. In that sense, it compares better with the extinction toward individual stars (in our Galaxy or others, e.g., M~31). On the other hand, as mentioned before, the dynamical range covered by the stellar reddening is of about two orders of magnitude, while it is only of only about one order of magnitude for the gas reddening. Ideally, to shed light on whether DIBs correlate better or worse with gas or stellar reddening, one would need to carry out a similar study to the one presented here in a system where gas reddening could be measured over about two orders of magnitude.
In any case, despite these caveats, this result suggests the possibility of using the DIBs as a tool for a first rough estimation of the reddening within a galaxy, in the absence of further information.
To do so, we model the relation between EW(DIB) versus\ $E(B-V)$ as a power law, following the formulation and nomenclature presented in \citet{Lan2015}, as follows:

\begin{equation}
    \label{eq:Power}
    \mathrm{EW(DIB)} = A \cdot E(B-V)^{\gamma}
,\end{equation}

where $\gamma$ and $A$ are the parameters to be fitted. The recovered functions for the different pairs of DIB-reddening are presented by the black dashed line in Fig. \ref{fig:DibGascomp}, and the corresponding $\gamma_t$ and $A_t$ are included in the upper-left corner of the corresponding graphic. 
For DIB$\lambda$5780 there is a minimal change in the slope between Fig. \ref{fig:DibGascomp} (a), the $E(B-V)_{\mathrm{Stellar}}$,  and (b), the $E(B-V)_{\mathrm{Gas}}$, even though in panel (a) all of our data lie above the fitted power law.
On the contrary, DIB$\lambda$6284 has its slope decreasing from $\gamma=0.7$, in relation to $E(B-V)_{\mathrm{Stellar}}$ (c), to $\gamma = 0.56$, in relation to $E(B-V)_{\mathrm{Gas}}$ (d). As such, the $E(B-V)_{\mathrm{Stellar}}$-DIB$\lambda$6284 compares better with the study by \citet{Lan2015}.

Panel (a) shows that the DIB$\lambda$5780 detected in our data is observed with a higher EW than reported in previous studies at similar $E(B-V)_{\mathrm{Stellar}}$. This difference could be explained by local ISM properties affecting the survival (or destruction) of the corresponding DIB carrier.
For example, it is known that the DIBs at $\lambda$5780, $\lambda$5797 and $\lambda$6284 present reduced EW in low metallicity environments (e.g., Large Magellanic Cloud, Small Magellanic Cloud) when they are compared with measurements in the Milky Way \citep{Welty2006, CoxMetal2006, Cox2007}. 
We can explore whether a dependence on the metallicity is present in the data by making two bins, one of studies only using data toward the Magellanic Clouds (MC) and one without (excluding \object{NGC 6240}). We can also fit a similar power law as before (Eq. \ref{eq:Power}). This result is also shown in Fig. \ref{fig:DibGascomp} and is presented by the dotted blue and dash-dotted red lines. Here we see that given a certain amount of attenuation, DIBs in a low metallicity environment (blue line) have systematically lower EWs than those detected in a higher metallicity environment (red line).

Building on this, from the galaxies presented in Fig. \ref{fig:DibGascomp}, the Milky Way and the Antennae Galaxy \citep[e.g.,][]{Lardo2015} have roughly about solar metallicity\footnote{Here, we assume $12+\log_{10}(\mathrm{O/H}) =8.69$ \citep{Asplund2021} as solar metallicity.}, while a slightly super-solar metallicity of value $12 + \log_{10}(\mathrm{O/H}) =8.85$ has been reported for NGC 6240 in the central region by \citet{Contini2012}.
The data points for M\,31 by \citet[inverted orange triangles in Fig.~\ref{fig:DibGascomp}]{Cordiner2011}, also with super-solar metallicity \citep{Sanders2012}, lie above the general function describing the relation between EW and reddening. 
We do not obtain a similar systematic EW abundance of the DIB$\lambda$6284 in panel (c) as for DIB$\lambda$5780 in panel (a), despite both DIBs being $\sigma$-types \citep{Fan2022} and highly correlated \citep[e.g.,][ determined $\rho$=0.96]{Friedman2011}. As discussed in Sec \ref{sec:Dibmaps} and displayed in Fig. \ref{fig:DIBmaps}, these DIBs are not all detected in the exact same tiles, and therefore they may well be located in ISM with different characteristics, including metallicity \citep[e.g.,][]{Herbig1993,Vos2011, Friedman2011}.
Altogether we see that gas reddening presents a stronger global correlation than stellar reddening for DIB$\lambda$5780, while the correlation is weaker for DIB$\lambda$6284. This result can be explained by several effects. First, due to the smaller dynamical range covered by the gas reddening, the result is prone to display spurious correlations and hide actual correlations. Second, the stellar and gas attenuation do not trace the same environment. Third, both DIBs are not detected in the same tiles. And lastly, due to the limited spectral and spatial resolution we are unable to disentangle various phases of the ISM. This result suggests the necessity for a more in-depth study that covers a range of physical and chemical properties in \object{NGC 6240} to shed light on components that play a role in the observed DIB and reddening relation.

\begin{figure}[h]
    \centering
    \resizebox{\hsize}{!}{\includegraphics{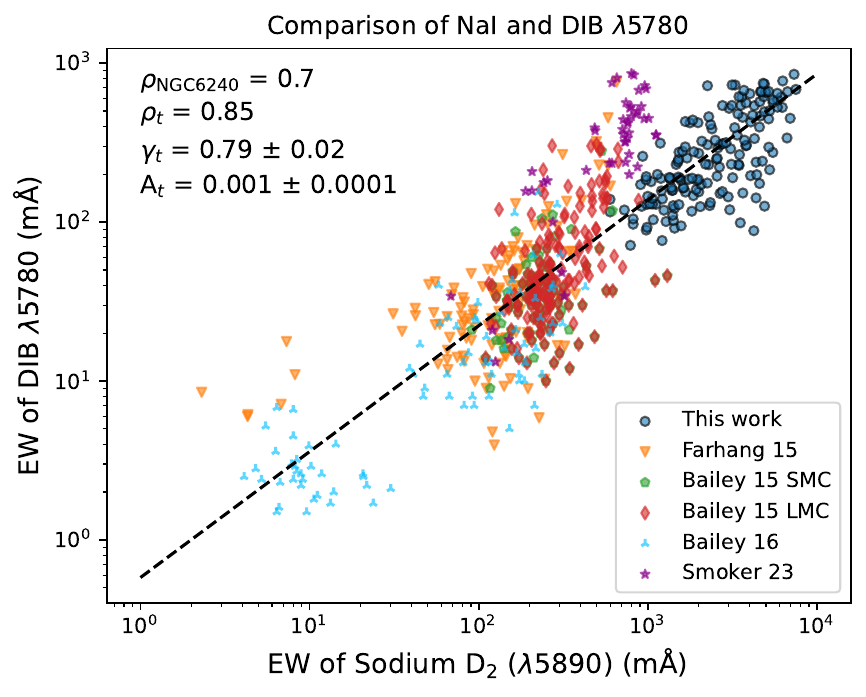}}\\
    \resizebox{\hsize}{!}{\includegraphics{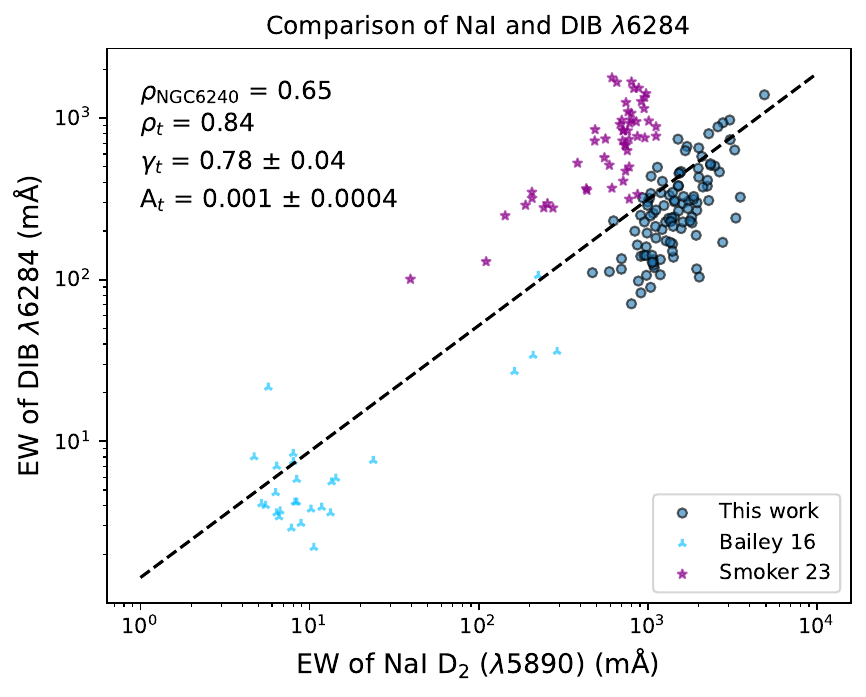}}
    \caption{Comparison of the equivalent widths of the \ion{Na}{i} D$_2$ line and the DIB$\lambda$5780 (top) and DIB$\lambda$6284 (bottom) features. The black dashed line is the power law fitted to all available data.}
    \label{fig:NaDibcomp}
\end{figure}

\subsection{DIBs and \ion{Na}{i} D}
In Figs. \ref{fig:DIBmaps} and \ref{fig:NAIDmap}, we present maps of both DIBs and the \ion{Na}{i} D$_2$ absorption feature. The spatial structure of these maps is relatively similar, even if  it is not entirely one-to-one.
In all maps, the equivalent widths are highest in the central area. Additionally, both the DIBs and the \ion{Na}{i} D are detected in the northern and southern dusty swathes, with lower values of EW than in the center, following the dust lane.
However, these maps differ east of the AGNs. In this area there is strong \ion{Na}{i} D detection but we do not observe the DIBs. The lack of detected DIBs in these tiles can be a purely observational effect, as the DIB being fully buried in a strong emission feature (i.e., the case of \oi$\lambda6300$ blended with DIB$\lambda 6284$). Another possibility could be an absence of DIB-producing material.
Results from previous studies have indeed shown that \ion{Na}{i} D and DIBs are correlated \citep{Sollerman2005, Welty2006, Farhang2015, Smoker2023}. However, this correlation solely suggests, if the conditions are such that a given DIB carrier can exist, that, for a given line of sight, both trace the amount of neutral ISM to the first order. Thus, it does not imply any physical similarities between DIB carriers and \ion{Na}{i} D \citep{Farhang2015, Ensor2017}. Moreover, studies in our Galaxy have already shown that atomic hydrogen can extend well beyond the area where DIBs can be detected (see Fig. 12 by \citealp{Puspitarini2015}). In the rest of this section, we  further explore the \ion{Na}{i} D-DIB correlation, on an extragalactic scale.
To do so, we followed a similar procedure as in Sec \ref{sec:DiscussionDIBebv}, and investigate the relationship in light of different studies. With respect to DIB$\lambda$5780, we added detections inside the Local Bubble \citep{Farhang2015, Bailey2016}, The Milky Way \citep{Smoker2023}, and Magellanic Clouds \citep{Bailey2015}. There are fewer studies available in the literature on the relationship between DIB$\lambda$6284 and \ion{Na}{i} D, which results in us only implementing results presented in \citet{Bailey2016} and \citet{Smoker2023}. Again, we fit a power law, as introduced in Eq. \ref{eq:Power}. Altogether, the results are presented in Fig. \ref{fig:NaDibcomp}. In \object{NGC6240} the DIBs are positively correlated with the \ion{Na}{i} D$_2$ ($\rho_\mathrm{\object{NGC6240},\lambda 5780} =0.70$ and $\rho_\mathrm{\object{NGC6240},\lambda 6284}= 0.65$).

Furthermore, we find that we only measured DIBs in tiles where EW(\ion{Na}{i} D$_2$) $\geq 1000$ m$\mathrm{\AA}$. This is only a small stretch of the entire EW(DIB)-EW(Na {\sc i} D2) relation that spans four orders of magnitude. In our study, the EW values, both for the DIBs as the \ion{Na}{i}, are relatively large with respect to what is measured in the Local Group. We can attribute this result to various effects.
These are: i) the substantial amount of dust and ISM present in a LIRG; ii) the inability to detect lower EWs due to our spectral resolution; and iii) and the limited spatial resolution.

When the correlation with respect to the previously mentioned studies ($\rho_t$) was calculated, we were able to obtain a relatively high Pearson correlation coefficient with values of $\sim 0.85$ for both the DIBs, as well as the exponent of the power law having value $\gamma_{t} \approx 0.79$.
Interestingly, the results reported by \citet{Smoker2023} show an offset with respect to our study, where they detect a lower Na I D equivalent width for a given EW(DIB). One possible cause of this discrepancy can be the different methodology to measure the EWs. To test this hypothesis we applied our algorithm to their DIB data and obtained a $\sim 30 \%$ reduction in the equivalent width. Although this adjustment brings the data points in closer relation to the fitted line, it does not fully account for the observed difference.

Another possible reason for this discrepancy could be the different (nature of) the areas sampled in each work. \citet{Smoker2023} utilized spectra from individual early-type stars, benefiting from higher spatial and spectral resolution. In contrast, our study involves a much lower spatial and spectral resolution. While \citet{Smoker2023} could measure DIBs toward single stars, our measurements are affected by the combination of multiple environments along a single sightline. 
A cookie-like configuration where, within a single tile, DIB carriers are confined to small patches embedded in a much larger volume, filled with atomic hydrogen, could explain the observed difference.

Even without knowing the carriers for these DIBs (or other DIBs), the results here suggest the possibility of using these features as astronomical tools. In our galaxy, the ratio of DIBs at $\lambda$5797/$\lambda$5780 has been suggested as a tracer of the ambient radiation field (\citealp[e.g.,][]{Krelowski1999, Sollerman2005, Vos2011, Ensor2017, Elyajouri2017}, although see \citealp{Lai2020} for a rebuttal). In a similar vein, we propose exploring the use of DIBs as tracers for the neutral gas content, particularly in cases where the \ion{Na}{I} D absorption feature becomes saturated \citep[as discussed in][and perhaps applicable to LIRGs]{Bailey2015} and no longer provides an accurate measurement of the column density and optical depth \citep{Puspitarini2014,Cazzoli2016}.

\subsection{DIBs and other phases of the ISM}

\begin{figure}[t]
    \centering
    \resizebox{\hsize}{!}{\includegraphics{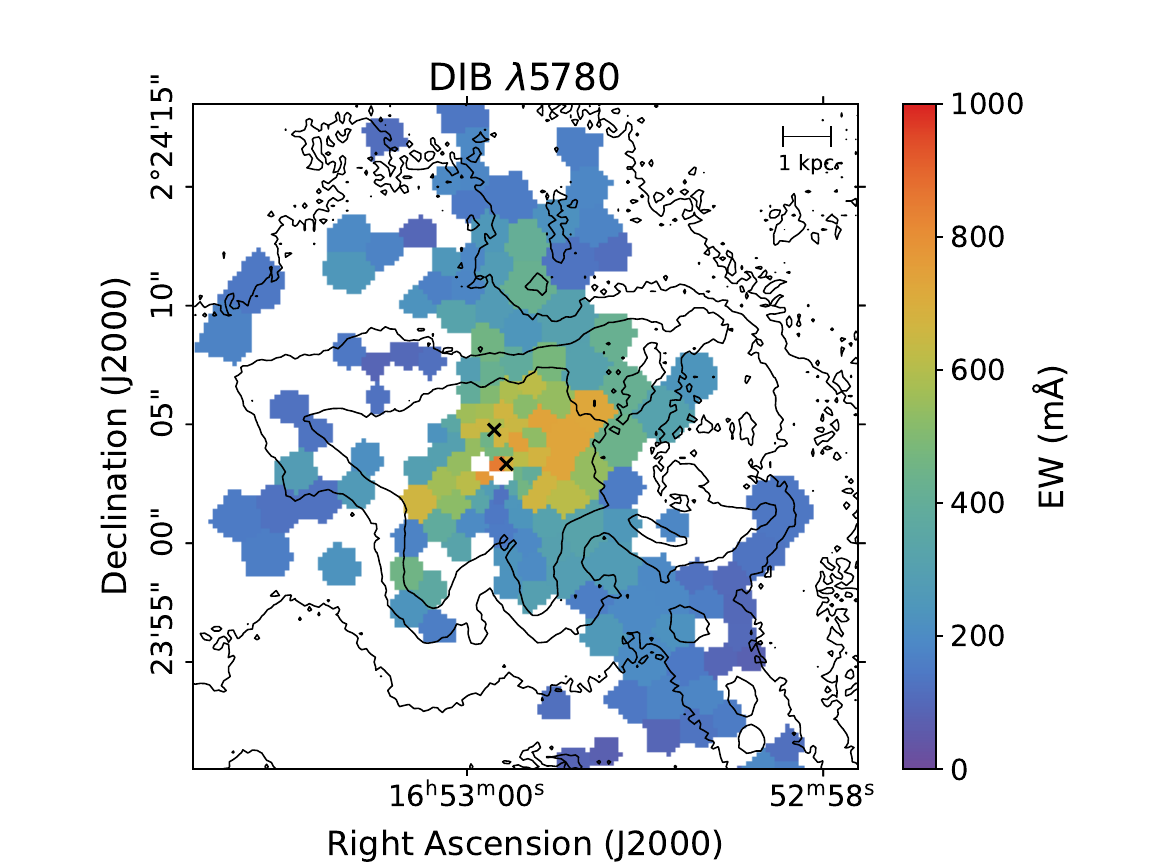}}\\
    \resizebox{\hsize}{!}{\includegraphics{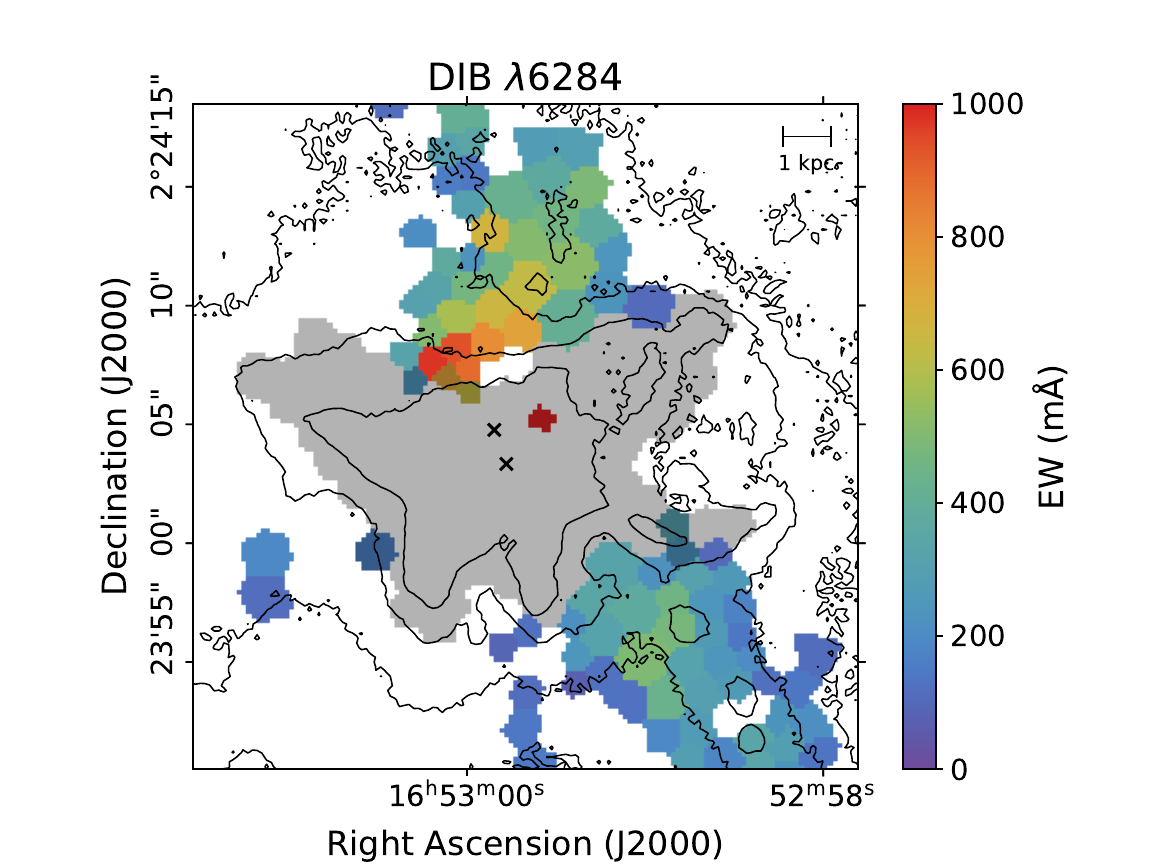}}
    
\caption{Map of DIB$\lambda$5780 (top) and $\lambda$6284 (bottom) near the AGNs of \object{NGC 6240}. These figures are all the exact same region that is marked by the white square in Fig. \ref{fig:datacube}. The black contours show the H$\alpha$ narrowband emission and the black crosses show the locations of the AGNs \citep{Komossa2003}.}
    \label{fig:CentreDib}
\end{figure}

In this section, we  explore the results of our work on DIB$\lambda$5780 and DIB$\lambda$6284 with other components of the ISM in \object{NGC 6240}; specifically, the H$_2$, H\,{\sc i}, and rotational CO features \citep[e.g.,][]{Treister2020, Fyhrie2021, Medling2021}. These studies only mapped the central area of \object{NGC 6240}, where only DIB$\lambda$5780 has been measured, which means we are unable to make any direct statements regarding DIB$\lambda$6284. As such, we limit the discussion to the DIB$\lambda$5780.

The top panel of Fig. \ref{fig:CentreDib} shows a zoom of DIB$\lambda$5780 nearby the AGNs in an area of $\sim 15 \mathrm{kpc} \times 15 \mathrm{kpc}$. We also present a similar zoom of DIB$\lambda$6284 in the bottom panel of Fig. \ref{fig:CentreDib} for consistency and as a reference for future work. We detect DIB$\lambda$5780 with large EW values in the vicinity and west of the two AGNs. This pattern is also observed in the radio continuum $\lambda20$~cm map  (Fig.\ 2a in \citealp{Colbert1994}), and the \ion{H}{i} absorption maps (Fig.\ 1 of \citealp{Beswick2001} and Fig.\ 6 of \citealp{Baan2007}). \citet{Baan2007} proposes this extended \ion{H}{i} absorption is due to the dust lane, which is supported by the observed relation between DIBs and $E(B-V)$ discussed in Sect.~\ref{sec:DiscussionDIBebv}. On the other hand, the $H_2$ and CO maps presented in Fig.\ 2 in \citet{Medling2021} and Fig.\ 12 in \citet{Max2005} show no strong emission west of the AGNs. This implies that DIB$\lambda$5780 present a spatial distribution more similar to the atomic than molecular spectral lines, where the latter traces colder ISM conditions \citep[also seen in][]{ MonrealIbero2018, Bailey2015, Lan2015}. This can be explained by the skin effect, which postulates that DIBs are formed on the edge of a cloud \citep{Herbig1995}.

\section{Conclusions}\label{sec:conclusion}
Here, we present the third of a series of experiments aiming at exploring the potential of highly efficient integral field spectrographs in extragalactic DIB research. For that purpose, we used archival MUSE data (in wide field mode) of a prototypical LIRG and system hosting two AGNs: \object{NGC 6240}. We extracted 488 spectra with a S/N>100 (which we based on the feature free spectral range of 5623 \r{A} - 5986 \r{A}) and searched for the well-known $\lambda$5780, $\lambda$5797, and $\lambda$6284 DIBs. We explored the (spatial) correlation between the detected DIBs and other characteristics of the ISM in the system, namely, the reddening, the \ion{Na}{i} D doublet, the molecular emission of H$_{2}$ and CO, and the $\lambda$21~cm \ion{H}{i} absorption feature. Our main findings can be summarized as follows:

\begin{enumerate}
    \item This system is, to our knowledge, the most distant one for which DIBs have ever been mapped,  at a distance of $\sim$100 Mpc. This is also the first time that DIB$\lambda$6284 has been mapped beyond the Local Group.
    
    \item We  detected  DIB$\lambda$5780 in 171 lines of sight, allowing us to map this feature over an total area of $\sim$ 79.96 kpc$^2$ at the center of the system. Likewise,  DIB$\lambda$6284 was detected in 109 lines of sight, over a total area of 59.78 kpc$^2$ in two distinguishable swathes north and south of the AGN, with the northern region covering an area of $\sim$21.22 kpc$^2$ and southern region covering an area of $\sim$31.41 kpc$^2$.
    
    \item The dust distribution in \object{NGC 6240} was mapped from the stellar attenuation, using the reddening of the stellar continuum and gas attenuation, based on the Balmer decrement, to the largest extent so far. These maps were composed of 451 and 459 lines of sight, respectively. Overall, with varying strength, the stellar attenuation is lower than the gas attenuation. The gas attenuation is in agreement with previous studies on the hydrogen emission features and both maps present a clearly defined 25 kpc-long dust lane extending from north to south.
    
     \item The observed DIBs are correlated with the dust distribution. In \object{NGC 6240}, DIB$\lambda$5780 and DIB$\lambda$6284 both have a stronger Pearson correlation with respect to $E(B-V)_{\mathrm{Stellar}}$ than for $E(B-V)_{\mathrm{Gas}}$, with $\rho_\mathrm{\object{NGC 6240}, \lambda 5780}$=0.76 and $\rho_\mathrm{\object{NGC 6240}, \lambda 6284}$=0.78 for $E(B-V)_{\mathrm{Stellar}}$. 

     \item With respect to other studies, we observe that DIB$\lambda$5780 has a larger Pearson correlation coefficient regarding the $E(B-V)_{\mathrm{Gas}}$  with respect to the $E(B-V)_{\mathrm{Stellar}}$. However, we are reluctant to interpret this purely mathematical result based on two grounds. First,  there is the relatively small dynamical range covered by the $E(B-V)_{\mathrm{Gas}}$ in our data ($\sim 1$dex). Second, the physical environment traced by $E(B-V)_{\mathrm{Stellar}}$  corresponds better to studies toward individual stars.
     
     \item We argue that both DIBs are generally better traced by the $E(B-V)_{\mathrm{Stellar}}$ than by the $E(B-V)_{\mathrm{Gas}}$, due to physical properties of the ISM. We determined a Pearson correlation coefficient of $\rho_t =0.77$ and a power law slope of $\gamma_t=0.81 \pm 0.03$ for DIB$\lambda$5780, along with correlation of $\rho_t =0.82$ and power law slope of $\gamma_t=0.7 \pm 0.03$ for DIB$\lambda$6240.
    
    \item There is a strong correlation between \ion{Na}{i} D and DIBs,  which we use to advocate the employment of DIBs as proxies for the amount of neutral gas in the limit where \ion{Na}{i} D is saturated. We support this statement with our mapping of the \ion{Na}{i} D absorption. Furthermore, we obtain a correlation of $\rho_t = 0.85$ and power law slope of $\gamma_t=0.79 \pm 0.02$ for EW(\ion{Na}{i}D)-EW(DIB$\lambda$5780) and correlation of $\rho_t = 0.88$ and power law slope of $\gamma_t=0.78 \pm 0.02$ for EW(\ion{Na}{i}D)-EW(DIB$\lambda$6284).

    \item None of the DIBs are spatially correlated with the molecular emission features H$_{2}$ and CO, however, they are better correlated with the \ion{H}{i} absorption feature.
\end{enumerate}

\begin{acknowledgements}
CvE would like to thank the EDIBLES consortium for the opportunity to present this work at a workshop. The authors would like to thank, and make a special mention in memory of, Prof. dr. H. V. J. Linnartz for his enthusiasm, support, inspiring words, and helpful discussions regarding this work. 
We also thank the referee for the valuable comments that have helped us to improve the first submitted version of this paper.
Based on observations collected at the European Southern Observatory under ESO
programmes 095.B-0482, 097.B-0588, and 099.B-0456.
This research has made use of the services of the ESO Science Archive Facility.
PMW gratefully acknowledges support by the BMBF from the ErUM program (project
VLT-BlueMUSE, grants 05A20BAB and 05A23BAC).

\end{acknowledgements}

\bibliographystyle{aa}
\bibliography{refs.bib}

\end{document}